\documentclass[]{aa}  

\usepackage{graphicx}
\usepackage{txfonts}
\usepackage{hyperref}
\usepackage{upgreek}
\usepackage{color}
\usepackage{comment}
\usepackage{mathtools}
\usepackage{ulem}
\usepackage{xspace}

\newcommand{\sectionname}{Sect.}
\newcommand{\dd}{\textrm{d}}
\newcommand{\di}{\textrm{i}}

\newcommand{\derivp} [2] {\frac {\partial #1 } {\partial #2} }

\newcommand{\deriv} [2] {\frac {\textrm{d} #1 } {\textrm{d} #2} }
\newcommand{\derivs} [2] {\frac {\textrm{d}^2 #1 } {\textrm{d} #2^2} }

\newcommand{\algn} [1] {
\begin{align} #1
\end{align}}

\let\originaleqref\eqref
\renewcommand{\eqref}{Eq.~\originaleqref}
\newcommand{\eq}[1] {Eq.\,(\ref{#1})}
\newcommand{\eqss}[2]{Eqs.~(\ref{#1})-(\ref{#2})}
\newcommand{\eqsduo}[2]{Eqs.~(\ref{#1}) and ~(\ref{#2})}
\makeatletter
\newcommand{\eqs}[1]{%
    Eqs.~(\ref{#1})\checknextarg}
\newcommand{\checknextarg}{\@ifnextchar\bgroup{\gobblenextarg}{}}
\newcommand{\gobblenextarg}[1]{\@ifnextchar\bgroup{, (\ref{#1})\gobblenextarg}{, and (\ref{#1})}}
\makeatother

\begin{document} 


   \title{Multi-cavity gravito-acoustic oscillation modes in stars}

   \subtitle{A general analytical resonance condition}
  \titlerunning{Multi-cavity oscillation modes} 
  
   \author{C. Pin\c con\inst{1,2} \and M. Takata\inst{3,4}
          }
          
  \authorrunning{C. Pin\c con and M. Takata}
   \institute{
        LERMA, Observatoire de Paris, Sorbonne Université, Université PSL, CNRS, 75014 Paris, France
        \and
        STAR Institute, Université de Liège, 19C Allée du 6 Août, B-4000 Liège, Belgium,
         \and
             Department of Astronomy, School of Science, The University of Tokyo, 7-3-1 Hongo, Bunkyo-ku, 113-0033 Tokyo, Japan
         \and
                LESIA, Observatoire de Paris, Université PSL, CNRS, Sorbonne Université, Université de Paris, Meudon, France
             }

   \date{\today}

  \abstract
   {Over recent decades, asteroseismology has proven to be a powerful method for probing stellar interiors. Analytical descriptions of the global oscillation modes, in combination with pulsation codes, have provided valuable help in processing and interpreting the large amount of seismic data collected, for instance, by space-borne missions CoRoT, {\it Kepler,} and TESS. These prior results have paved the way to more in-depth analyses of the oscillation spectra of stars in order to delve into subtle properties of their interiors. This purpose conversely requires innovative theoretical descriptions of stellar oscillations.}
  {In this paper, we aim to  analytically express the resonance condition of the adiabatic oscillation modes of spherical stars in a very general way that is applicable at different evolutionary stages.}
   {In the present formulation, a star is represented as an acoustic interferometer composed of a multitude of resonant cavities where waves can propagate and the short-wavelength JWKB approximation is met. Each cavity is separated from the adjacent ones by barriers, which corresponds to regions either where waves are evanescent or where the JWKB approximation fails. Each barrier is associated with a reflection and transmission coefficient. The stationary modes are then computed using two different physical representations:\ 1)  studying the infinite-time reflections and transmissions of a wave energy ray through the ensemble of cavities or 2) solving the linear boundary value problem using the progressive matching of the wave function from one barrier to the adjacent one between the core and surface.}
   {Both physical pictures provide the same resonance condition, which ultimately turns out to depend on a number of parameters: the reflection and transmission phase lags introduced by each barrier, the coupling factor associated with each barrier, and the wave number integral over each resonant cavity. Using such a formulation, we can retrieve, in a practical way, the usual forms derived in previous works in the case of mixed modes with two or three cavities coupled though evanescent barriers, low- and large-amplitude glitches, and the simultaneous presence of evanescent regions and glitches.}
   {The resonance condition obtained in this work provides a new tool that is useful in predicting the oscillation spectra of stars and interpret seismic observations at different evolutionary stages in a simple way. Practical applications require more detailed analyses to make the link between the reflection-transmission parameters and the internal structure. These aspects will be the subject of a future paper.}

   \keywords{asteroseismology -- stars: oscillations -- methods: analytical}

   \maketitle


%
\section{Introduction}
\label{introduction}

Mechanical forcing, turbulent motions, or thermal instabilities can perturb the equilibrium state of stars and generate internal waves \citep[e.g.,][and references therein]{Samadi2015}. These waves can propagate back and forth several times between the center and surface of stars and a resulting positive interference can give rise to global resonant modes, the oscillation frequencies of which directly depend on the stellar interior properties. The study of the oscillation power spectrum of these modes (asteroseismology) can, in turn, provide us with precious information on the stellar structure. Over recent decades, the high-quality seismic data first provided for the Sun by the spacecraft SoHO \citep{Domingo1995} and ground-based telescope networks \citep[e.g., GONG and BiSON projects,][]{Chaplin1997,Leibacher1998} as well as subsequent data gathered  for thousands of distant stars by space-borne missions CoRoT, {\it Kepler,} and TESS \citep{Baglin2006b,Borucki2010,Ricker2015}, have indeed brought stringent constraints on the stellar dynamics and evolution \citep[e.g.,][]{Chaplin2013,Noels2016b,Mosser2016,JCD2020}. The exploitation of such an amount of information and its physical interpretation have greatly relied on analytical descriptions of the linear stellar oscillations. In particular, theoretical expressions of the eigenfrequency patterns appeared to be crucial for extracting the prevailing features of the observed oscillation power spectra, defining seismic indicators that are relevant of the stellar structure, and enabling statistical studies on large samples of stars via automated methods \citep[e.g., see][for recent examples]{Farnir2019,Appourchaux2020,Gehan2021}.

Most of the available analytical descriptions of the linear stellar oscillations are based on asymptotic methods within the JWKB approximation. This approximation assumes that the wavelength is much smaller than the variation scale height of the background state almost everywhere in stars  \citep[e.g.,][]{Olver1975,Gough2007}. For example, asymptotic analyses predict that the high-frequency acoustic modes, the propagation cavity of which is mostly located in the uppermost layers of stars, are nearly evenly spaced in frequency, with a frequency spacing (or large frequency separation) directly linked to the mean density of stars. In contrast, the low-frequency gravity modes, which mostly propagate in the inner layers of stars, are asymptotically expected to be nearly evenly spaced in period, with a period spacing related to the stellar core density \citep{Vandakurov1968,Tassoul1968,Smeyers1968,Shibahashi1979,Tassoul1980}. This leading-order description of the mode frequency distribution is nevertheless insufficient for reproducing all the diversity and details observed in the high-quality oscillation spectra of the Sun and thousands of other stars.

First, in low-mass main sequence stars, the deviation of the observed acoustic mode frequencies from the nearly regular asymptotic pattern was shown to take the form of a low-amplitude sinusoidal-like signal \citep[e.g.,][]{Hill1986}. The origin of these small frequency perturbations is attributed to sharp structural variations in some regions of stars where the JWKB approximation fails, the so-called glitches \citep[][]{Vorontsov1988,Gough1988,Gough1990}. For instance, glitches are expected at the interface between convective and radiative zones where discontinuities in the temperature gradient, in its derivatives, or in the molecular weight can occur, depending on the mixing processes at work \citep[e.g.,][]{Gough1993,Roxburgh1994,Monteiro1994,Audard1994,Monteiro2000}, as well as in the helium ionization zone where the first adiabatic index abruptly varies \citep[e.g.][]{Gough1995,Gough2002,Basu2004,Houdek2007}. Most prior works have assumed that the amplitude of these acoustic glitches is small enough so that the induced deviation from the asymptotic frequency pattern can be analytically modeled using first-order perturbation methods based on the variational principle. These developments have permitted in-depth studies of acoustic glitches in the Sun and brought about stringent constraints on the position of the base of the convective envelope, on the extent of the overshooting region beneath \citep[e.g.,][]{Basu1994,JCD1995,Basu1997}, and on the location of the ionization zone and the surface helium abundance \citep[e.g.,][]{Vorontsov1991,Perez1994,Monteiro2005}. The exquisite store of seismic data collected by space-borne missions CoRoT and {\it Kepler} subsequently enabled similar studies in dozens of other distant main-sequence stars \citep[e.g.,][]{Mazumdar2012,Lebreton2012,Mazumdar2014,Verma2017,Farnir2020b}. All the seismic information extracted from acoustic glitches currently stand for one of the most important sources of constraints for stellar modeling \citep[e.g.,][]{Verma2019,Farnir2020}.

Furthermore, in intermediate-mass main sequence stars, the frequency pattern of the observed gravity modes can also be affected by glitches. In these stars, the expansion and recession of the convective core create a gradient in the mean molecular weight at its upper boundary, resulting in a large discontinuity in the Brunt-Väisälä frequency \citep[][]{Berthomieu1988,Provost1990}. Such large-amplitude glitches locally induce an important wave reflection and result in an unequal distribution of the mode energy on both sides of the glitch: this is referred to as the mode-trapping phenomenon. Unlike low-amplitude glitches in the Sun, the deviations from the asymptotic frequency pattern of gravity modes that are induced by such large-amplitude glitches cannot be treated as small perturbations. In this situation, the eigenfrequencies of gravity modes can be better analytically expressed using the matching of the asymptotic solutions on each side of the glitch \citep[e.g.,][]{Miglio2008b,Miglio2008}. The period spacing between adjacent gravity modes predicted by such models significantly differ from the uniform asymptotic value predicted in the absence of discontinuity, which offers an interesting potential to constrain the chemical mixing at the edge of convective cores in these stars \citep[e.g.,][]{Degroote2010}. The frequency pattern of gravity modes in white dwarfs can similarly be affected by compositional layering and therefore can similarly provide information on the internal structure and transport processes in these advanced evolutionary stages \citep[e.g.,][]{Brassard1992,Kawaler1994}.

Finally, in post-main sequence stars, the density contrast between the core and surface is so large that the oscillation modes can propagate both in an inner cavity, where they behave as gravity modes, and in an outer cavity, where they behave as acoustic modes: these are the so-called mixed modes \citep{Scuflaire1974,Aizenman1977}. Both cavities are coupled by an intermediate region where the modes are evanescent \citep[e.g.,][for a review]{Hekker2017}. Neglecting the possible effect of internal glitches, the asymptotic frequency pattern of mixed modes presents the characteristics of both the acoustic and gravity modes spectra \citep{Shibahashi1979,Tassoul1980,Takata2016a,Loi2020}. The asymptotic analyses of mixed modes appeared essential to analyze the large amount of seismic data collected by the satellite CoRoT and {\it Kepler} for evolved stars and represent a solid theoretical basis for interpreting these observations in terms of internal structure \citep[e.g.,][]{Mosser2018}. The study of mixed modes put stringent constraints for instance on the nuclear-burning state \citep[e.g.,][]{Montalban2010,Mosser2014,Vrard2016}, the core rotation
\citep[e.g.,][]{Goupil2013,Gehan2018,Deheuvels2020}, and the mid-layer structure of these stars \citep[e.g.,][]{Mosser2017b,Hekker2018,Khan2018,Pincon2019,Pincon2020}, as well as on the amount of core overshooting on the main sequence \citep[e.g.,][]{Montalban2013,Noll2021}. Delving into more details of the mixed mode oscillation spectra, acoustic glitches associated with the helium ionization zone could also be detected \citep{Vrard2015,Dreau2020}. Other seismic signatures, still not observed, have also been theoretically predicted. On the one hand, \cite{Cunha2015,Cunha2019} proposed a more complex description of mixed modes including the presence of buoyancy glitches, that is, the sharp gradient in the molecular weight induced by the migration of the base of the convective zone during the first dredge-up. On the other hand, \cite{Deheuvels2018} investigated the properties of mixed modes during the ignition of the helium burning (i.e., helium sub-flashes). During this phase, the temporary helium-burning shell is convective and mixed modes become evanescent inside. The inner propagation cavity is thus split into two parts. Mixed modes can thus propagate in three cavities separated from each other by two evanescent coupling regions. Both previous internal features were shown to produce remarkable seismic signatures in the mixed mode spectra, with promising probing potentials.

All the previous examples demonstrate the diversity of potential resonance configurations throughout the Hertzsprung-Russel diagram and the large amount of associated analytical descriptions. Despite this diversity, the effect of low- and large-amplitude glitches as well as evanescent regions on the mode frequencies results from a similar physical phenomenon. Indeed, either near sharp gradients or close to the boundaries between evanescent regions and resonant cavities (i.e., close to turning points where the radial wave number vanishes), the variation scale height of the background state is smaller than the local oscillation wavelength, the JWKB approximation locally fails, and an incident wave energy ray is partially reflected and transmitted \citep[see, e.g., Appendix~C of][for a simple example]{Pincon2020}. Based on basic linear wave principles, \cite{Takata2016b} considered such a physical picture and reformulated the mixed mode frequency pattern of red giant stars in a very general way by describing the evanescent region as a simple barrier associated with a wave reflection coefficient. \cite{Pincon2019b} adapted the same picture, while adding the presence of one glitch, which had been actually already anticipated by \cite{Roxburgh2001}. In this paper, we aim to extend these previous representations to a multitude of resonant cavities and barriers  and to obtain a generalized formulation for the resonance condition of oscillation modes in spherical stars that is applicable to any configurations and evolutionary states in a practical way. 

The paper is organized as follows. In \sectionname{}~\ref{setting}, we introduce the theoretical background about gravito-acoustic oscillations in stars and the modeling of the wave transmission-reflection problem. This introductory material is subsequently applied to obtain a very general expression of the resonance condition following two complementary pictures in Sects.~\ref{infinite reflection} and \ref{BV picture}. In addition, the distribution of the mode energy and the mode amplitudes among the different resonant cavities are addressed in \sectionname{}~\ref{mode energy and amplitude}. As a first illustration, the resulting resonance expression is then applied on simple usual oscillation configurations in \sectionname{}~\ref{simple cases} and the compatibility with previous formulations is discussed. We present our concluding remarks in \sectionname{}~\ref{conclusion}.

%
\section{Setting the stage}
\label{setting}

In this first section, we introduce the theoretical background and set the main physical description of the stellar oscillation modes used in the present paper.

\subsection{General theoretical framework}

In this work, we focus on the linear asymptotic and adiabatic global standing modes of spherical stars. In other words, we make the three following main assumptions on the oscillations. 

First, we assume that the oscillations are small-amplitude perturbations of the stellar equilibrium state (i.e., within the linear approximation) and that there is at least one region where they can propagate as progressive waves and where the short-wavelength JWKB approximation is met  (i.e., within the asymptotic limit). Such a region is referred to as a resonant cavity. 

Second, we assume that the waves are generated at a time $t=-\infty$, and we neglect non-adiabatic effects during their propagation (i.e., within the adiabatic limit). We therefore focus on stationary modes that persist indefinitely over time with a time dependence in $e^{-\di\sigma t}$, where $\di$ is the imaginary unit and $\sigma$ is the angular oscillation frequency. This seems reasonable in a first step since, otherwise, excited progressive waves would be rapidly damped and could not constructively interfere to form eigenmodes.

Third, we neglect the effects of stellar rotation and internal magnetic effects on the oscillations; in other words, there is no preferential axis from the point of view of the waves. Owing to the spherical symmetry, the perturbations associated with the oscillations are separable into angular and radial parts, and the angular part is represented by the orthonormal set of scalar and vector spherical harmonics, $Y_\ell^m$, in the stellar frame, with angular degrees, $\ell,$ and azimuthal numbers, $m$ \citep[e.g.,][]{Ledoux1958,Unno1989}. In the following, we explicitly focus on a $(\ell,m)$ harmonic; the Eulerian perturbation of pressure, $p^\prime$, and the oscillation displacement field, $\vec{\xi}$, are thus expressed as:
\algn{
p^\prime(\vec{r},t)&=\tilde{p}^\prime(r)~Y_\ell^m(\theta,s)~ e^{-\di\sigma t} \label{p pert},\\
\vec{\xi}(\vec{r},t)&=  \left[\tilde{\xi}_{r}(r)~Y_\ell^m(\theta,s) ~\vec{e}_r + \tilde{\xi}_{h}(r) ~r \vec{\nabla} Y_\ell^m(\theta,s) \right] e^{-\di\sigma t} \label{xi pert}\; ,
}
where $(r,\theta,s)$ are the spherical coordinates in the stellar frame, $\vec{e}_r$ is the radial unit vector, $\vec{\nabla}$ is the gradient operator, and \smash{$\tilde{\xi}_{r}$ and $\tilde{\xi}_{h}$} are the radial and poloidal components of the mode displacement, respectively.

The present theoretical framework is therefore appropriate to investigate the frequency pattern of eigenmodes in the slow rotator, low magnetic field, and adiabatic limits. We note that the study of the small deviations from this leading-order pattern that can be induced by rotational, magnetic, and non-adiabatic effects is theoretically tractable in a subsequent step using perturbative methods \citep[e.g.,][]{Ledoux1951,Dziembowski1977b,Dziembowski1984,Pincon2021}, which is beyond the scope of this paper. Once the framework is set, we introduce in the following the basics on the propagation of gravito-acoustic waves that are useful for our purpose while distinguishing two kinds of regions: first, the resonant cavities where progressive waves can propagate and where the JWKB approximation is met; and second, the regions where waves are evanescent or where the JWKB hypothesis fails at some point (i.e., in the vicinity of turning points or near sharp structural variations). These regions are referred to as barriers in the following.

\subsection{Wave propagation in resonant cavities}

\subsubsection{Adiabatic linear wave equation}

Assuming an adiabatic equation of state, the continuity, momentum and Poisson equations linearized around the hydrostatic equilibrium of stars results in a fourth-order differential system with respect to radius for the radial wave displacement, the Eulerian pressure perturbation, the Eulerian perturbation of the gravitational potential, and the radial derivative of this latter \citep[e.g.,][]{Ledoux1958}. Nevertheless, within the short-wavelength JWKB assumption, the effect of the perturbation of the gravitational potential on the wave displacement field is known to be negligible at leading order. This is the so-called Cowling approximation \citep{Cowling1941}. It relies on the fact that because of the large number of radial nodes in the vicinity of the considered layer, the effect of the small-scale density perturbations integrated over the whole stellar volume is negligible on the local gravitational acceleration according to the Poisson equation \citep[see, e.g., Appendix A of][for a scaling-based justification]{Pincon2020}.
The validity of the Cowling approximation for any angular degree, $\ell,$ inside the resonant cavities thus permits us to reduce the adiabatic linear wave equation to the second order, which has the advantage of being analytically tractable using the usual asymptotic methods \citep[e.g.,][]{Olver1975}.

In these considerations, the formulation of \cite{Shibahashi1979} within the Cowling approximation can be used to describe the leading-order behavior of the oscillations inside resonant cavities. According to the work of \cite{Shibahashi1979}, the wave equation can be expressed as
\algn{
\derivs{\Psi}{r} + \left[\mathcal{K}_r^2-\frac{M(r)}{H_p^2}\right] \Psi = 0 \; ,
\label{wave equation}
}
where the wave function and the squared local asymptotic radial wave number are respectively defined as
\algn{\Psi &=\rho^{1/2} c r\left| 1-\frac{S_\ell^2 }{\sigma^2}\right| ^{-1/2}  \tilde{\xi}_{r} ~e^{-\di\sigma t}\; ,
\label{def v}\\
\mathcal{K}_r^2 &= \frac{\sigma^2}{c^2}\left(\frac{N^2}{\sigma^2} -1\right) \left(\frac{S_\ell^2}{\sigma^2} -1\right) \; .
\label{K_r JWKB}
}
In these equations, $\rho$ is the equilibirum density, $c^2=\Gamma_1 p/\rho$ is the squared sound speed, with $\Gamma_1$ the first adiabatic index and $p$ the equilibrium pressure, $H_p$ is the pressure scale height, and $S_\ell$ and $N$ are the Lamb and Brunt-Väisälä frequencies, respectively, the expressions of which are provided by
\algn{
S_\ell^2&=\frac{\ell(\ell+1) ~c^2}{r^2}\\
N^2&=\frac{g}{r}\left (\frac{1}{\Gamma_1}\deriv{\ln p}{\ln r}-\deriv{\ln \rho}{\ln r} \right) \; ,
}
where $g$ is the gravitational acceleration. Finally, the $M(r)$ function in \eq{wave equation} is a radial function that depends on the variation of the equilibrium structure (see \appendixname{}~\ref{Shiba} for an expression). 

As mentioned earlier, the second-order wave equation in \eq{wave equation} can be considered as a very good approximation inside regions where the JWKB approximation is met. By definition, these regions are far enough away from sharp variations in the stellar structure (i.e., relative  to the local wavelength) in such a way that the $M$ function in \eq{wave equation} remains on the order of unity at most inside these regions\footnote{The $M(r)$ function in \eq{wave equation} is singular at turning points where $\smash{\sigma=S_\ell}$. This singularity is however not an issue for the discussion since it is not physical and can be easily removed by changing the dependent variable appropriately, as recalled in \appendixname{}~\ref{Shiba}.}. We therefore understand that the resonant cavities correspond to regions where $\mathcal{K}_r^2>0$ that are located between consecutive turning points (i.e., where $\mathcal{K}_r^2=0$ for $\sigma^2=N^2$ or $\sigma^2=S_\ell^2$) and sharp gradients, but far enough away from these latter layers for the JWKB approximation to be met (i.e., $\mathcal{K}_r^2\gg1/H_p^2$). According to \eq{K_r JWKB}, this is the case where \smash{$\sigma^2 \ll (S_\ell^2~\mbox{and} ~N^2)$} or $\sigma^2 \gg (S_\ell^2~\mbox{and} ~N^2)$, which corresponds to low-frequency gravity-dominated waves or high-frequency pressure-dominated waves, respectively.

From a general point of view, one particular solution of the wave equation in \eq{wave equation}, denoted by $\psi$, can be formally written in a plane wave form  as
\algn{
\psi(r)=e^{\di \phi(r)} \; ,
\label{exp form}
}
where $\phi(r)$ is a complex phase function. It is easy to check that its complex conjugate, $\psi^\star$, is also a solution. Moreover, their Wronskian $\mathcal{W}_\psi$ is equal to 
\algn{
\mathcal{W}_\psi=\left(\psi^\star\deriv{\psi}{r}-\psi\deriv{\psi^\star}{r} \right)= 2 \di \mathcal{R}_{\rm e}\left[\deriv{\phi}{r}\right] \left| \psi \right|^2\; ,
\label{wronskian}
}
where $\mathcal{R}_{\rm e}[\cdot]$ denotes the real part. Therefore, $\psi$ and $\psi^\star$ are linearly independent and thus form a vectorial basis for the solutions of the wave equation if and only if $\mathcal{R}_{\rm e}[\dd \phi/\dd r]\ne 0$. This condition is met inside resonant cavities, since $\mathcal{K}_r^2>0$. It is also worth noting by differentiating \eq{wronskian} with respect to $r$ and using \eq{wave equation} that their Wronskian $\mathcal{W}_\psi$ is conserved in resonant cavities, as expected from the Liouville formula.

\subsubsection{Asymptotic form of the wave solution}

In resonant cavities, we denote \smash{$\mathcal{K}_r=\sqrt{\mathcal{K}_r^2}$}. We assume that the phase $\phi$ in \eq{exp form} varies on a length scale on the order of $1/\mathcal{K}_r$, which is supposed to be much smaller than the variation scale height of the medium on the order of $H_p$.
Under these considerations, injecting \eq{exp form} in \eq{wave equation}, we can show that the $\phi$ function in the cavity is equal at leading order in $\mathcal{K}_r$, up to addition by a constant, to \citep[e.g.,][]{Gough2007}
\algn{
\phi(r;\bar{r})\approx\frac{\di}{2}\ln \mathcal{K}_r + \varphi(r;\bar{r}) \; ,
\label{phi JWKB}
}
where
\algn{
\varphi(r;\bar{r}) = \int_{\bar{r}}^r \mathcal{K}_r \dd r \; ,
\label{phase}
}
with $\bar{r}$ an arbitrary reference point in the considered cavity. In the latter equation, we have chosen as a convention the positive branch in \eq{phi JWKB} such as $(d \varphi / dr)=\mathcal{K}_r>0$ and the $\psi$ function is associated with the progressive component of the wave function $\Psi$ whose phase travels upward (given the temporal dependence on $e^{-\di\sigma t}$). Therefore, in resonant cavities, the $\psi$ function merely reads at leading order 
\algn{
\psi(r;\bar{r})\approx\frac{1}{\sqrt{\mathcal{K}_r}} e^{\di \varphi(r;\bar{r})} \; ,
\label{psi sol}
}
which corresponds to a plane wave with a slowly varying amplitude. Given \eq{phi JWKB}, the Wronskian of the two basis solutions in \eq{wronskian} is equal at leading order to \smash{$\mathcal{W}_\psi\approx 2 \di \mathcal{K}_r  /\mathcal{K}_r = 2\di$}, which is constant as expected.

\subsubsection{Wave energy luminosity}

Physically speaking, the propagation of gravito-avoustic waves is not only described by the propagation of their phases, but also by the propagation of their energy. To analyze this, we find convenient to consider the radial wave energy luminosity.
Within the Cowling approximation (resulting from the short-wavelength hypothesis), it is defined at leading order by the integrated quantity at time $t$ and radius $r$ \citep[e.g.,][]{Lighthill1978,Unno1989}:
\algn{
\overline{\mathcal{L}_{{\rm w}}} \approx \frac{1}{T} \int_{t-T/2}^{t+T/2}\left(\iint_{\theta,s}\mathcal{R}_{\rm e} \left[ p^\prime (\vec{r},\tau)\right] \mathcal{R}_{\rm e}\left[ \varv_r (\vec{r},\tau) \right]r^2 \sin\theta \dd \theta \dd s \right)\dd \tau \; ,
\label{radial flux}
}
where $\varv_r$ is the Eulerian perturbation of radial velocity and $T$ is the oscillation period.
By recalling that the wave velocity field is given by $\vec{\varv}=(\partial \vec{\xi}/\partial t)$ and by expressing within the considered framework its radial component as
\algn{
\varv_r(\vec{r},t)=\tilde{\varv}_r (r) ~Y_\ell^m(\theta,s) ~e^{-\di\sigma t} \label{v_r pert}\; ,
}
we obtain the simple relation
\algn{
\tilde{\varv}_r(r)=-\di \sigma \tilde{\xi}_r(r) \; .
\label{v_r xi}
}
In order to express \eq{radial flux} further within the JWKB approximation, we then write the wave function $\Psi$ in a general way as the linear combination of $\psi$ and $\psi^\star$, that is,
\algn{
\Psi(r,t)\approx e^{-\di \sigma t} \left[a_{\rm p}~ \psi (r,\bar{r})+ a_{\rm r} ~\psi^\star (r;\bar{r}) \right]\; ,
\label{sol general}
}
where $a_{\rm p}$ and $a_{\rm r}$ are two complex constants representing the amplitudes of the progressive and regressive components, respectively. Using \eqs{def v}{K_r JWKB}{psi sol}{v_r xi}, the (complex) radial part of the velocity in resonant cavities is therefore equal to:
\algn{
\tilde{\varv}_r\approx -\di \left(\frac{\sigma}{\rho r^2 c}\right)^{1/2} \left| \frac{S_\ell^2-\sigma^2}{N^2-\sigma^2}\right| ^{1/4}  \left( a_{\rm p}~e^{i\varphi} + a_{\rm r}~ e^{-i\varphi}\right)\; .
\label{v_r}
}
To obtain the Eulerian perturbation of pressure, we can then use the relations derived by \cite{Shibahashi1979}, and given in \eqsduo{var w}{p by xi},~while neglecting at leading order the variations of the structure equilibrium compared to that of the wave phase, which provides
\algn{
\tilde{p}^\prime \approx{\rm sgn}  (S_\ell^2-\sigma^2)  \di \left(\frac{\sigma \rho c}{r^2} \right)^{1/2} \left|\frac{N^2-\sigma^2}{S_\ell^2-\sigma^2}\right| ^{1/4} \left( a_{\rm p}~e^{i\varphi} - a_{\rm r}~ e^{-i\varphi}\right)\; ,
\label{p^prime}
}
where sgn() is the sign function.
Therefore, using \eqs{p pert}{v_r pert}{v_r}{p^prime}, and taking advantage of the time dependence on $e^{-\di \sigma t}$ and the orthonormality of the spherical harmonics, it is straightforward to show that the radial wave energy luminosity in \eq{radial flux} finally reads
 \algn{
 \overline{ \mathcal{L}_{{\rm w}}} \approx {\rm sgn}  \left(\sigma^2-S_\ell^2\right) \frac{ \sigma }{2} (|a_{\rm p}|^2-|a_{\rm r}|^2) \; .
 \label{mean flux}
 }
First, we check that the radial wave energy luminosity does not depend on $r$ and $t$ within the hypothesis of adiabatic oscillations, as expected. Second, we see that it is equal to the sum of two distinct parts that are proportional to $|a_{\rm p}|^2$ and $|a_{\rm r}|^2$, which result, respectively, from the progressive and regressive components\footnote{In general, the oscillations are represented by a linear combination of an infinite number of $(\ell,m)$ harmonics. Given the orthonormality of the spherical harmonics, it is straightforward to show that the total mean wave energy luminosity is equal to the sum of each harmonic contribution, and that each harmonic contribution has the same form as \eq{mean flux}.}. Third, we see that in general, the mean wave energy flux is provided by
 \algn{
 \overline{ \mathcal{F}_{{\rm w}}} (r) \equiv \frac{\overline{\mathcal{L}_{\rm w}}}{4\pi r^2}\approx \frac{1}{16\pi } \left( \tilde{p}^\prime \tilde{\varv}_r^\star+\tilde{p}^\prime{}^\star \tilde{\varv}_r\right) \; .
 }
In the case where there is only one progressive or one regressive component (i.e., $a_{\rm r}=0$ or $a_{\rm p}=0$, respectively), we retrieve the well-known formula for plane waves given by \smash{$\overline{\mathcal{F}}_{\rm w}=\tilde{p}^\prime \tilde{\varv}_r^\star/8\pi$.}
Finally, it is worth mentioning that the Wronskian for the wave function $\Psi$ is equal to 
 \algn{
 \mathcal{W}_\Psi = \left(\Psi^\star\deriv{\Psi}{r}-\Psi\deriv{\Psi^\star}{r} \right) = (|a_{\rm p}|^2-|a_{\rm r}|^2) \mathcal{W}_\psi \; .
 }
At leading order in the resonant cavities, $\mathcal{W}_\psi\approx 2\di $, and we see that
the Wronskian of the wave function $\Psi$ is thus directly related to the radial wave energy luminosity, which explains its conservation within the adiabatic limit.

Another important point is the presence of the term \smash{${\rm sgn} (\sigma^2-S_\ell^2)$} in \eq{mean flux}. This term is actually related to the direction of the propagation of the wave energy. Indeed, the radial group and phase velocities in resonant cavities, $\varv_{\rm g}$ and $\varv_{\varphi}$, are equal, according to \eq{K_r JWKB}, to:
\algn{
\varv_{\varphi}=\pm \frac{\sigma}{\mathcal{K}_r} ~~~~{\rm and}~~~~\varv_{\rm g}=\derivp{\sigma}{\mathcal{K}_r} =\frac{c^2 \mathcal{K}_r^2 \sigma^2}{\left( \sigma^4-N^2S_\ell^2\right)}\varv_{\varphi} \; ,
}
where the plus and minus signs correspond to the progressive and regressive components, respectively. Therefore, two cases have to be distinguished. On the one hand, in the case of pressure-dominated oscillations, we have \smash{$\sigma^2 \gg (S_\ell^2~\mbox{and}~N^2)$} so that $\varv_{\rm g}$ and $\varv_\varphi$ have the same sign. As a consequence, the wave energy propagates in the same direction as the phase wavefront. Therefore, the progressive (regressive) component is associated with a positive (negative) mean wave energy flux propagating upward (downward). On the other hand, in the case of gravity-dominated oscillations, we have \smash{$\sigma^2 \ll (S_\ell^2~\mbox{and}~N^2)$} so that $\varv_{\rm g}$ and $\varv_\varphi$ have opposite signs. The wave energy propagates in the opposite direction of the phase wavefront. In this case, the progressive (regressive) component is associated with a negative (positive) mean wave energy flux propagating downward (upward). This behavior is in agreement with the term \smash{${\rm sgn} (\sigma^2-S_\ell^2)$} in \eq{mean flux}.

\subsection{Barriers and basic wave reflection-transmission problem}
 \label{barrier}
 
There are some regions in stars where the waves are evanescent (i.e., where $\mathcal{K}_r^2<0$) or where the JWKB approximation is not met, as in the vicinity of sharp variations in the medium (e.g., discontinuity in the chemical composition, thin ionization region) or turning-points (i.e., where $\mathcal{K}_r^2 = 0$). These are the so-called barriers.
While in the case of one infinite-length cavity a purely progressive or regressive solution can exist (i.e., as in \eq{sol general} with $a_{\rm r}=0$ or $a_{\rm p}=0$), this does not hold true in the presence of one single barrier. Indeed, an incident wave coming from a first cavity is partially reflected back near the barrier and partially transmitted toward a second cavity on the other side of the barrier. It is thus necessary to consider the reflected and transmitted waves as well, the amplitudes of which depend both on the amplitude of the incident wave and on the properties of the barrier, and they have to satisfy the conservation of the wave energy flux. In general, this physical problem can be described by reflection and transmission coefficients, as already proposed by \cite{Roxburgh2001} or \cite{Takata2016b} in the context of stellar pulsations.

To describe the wave reflection-transmission problem, we considered a first cavity $C_n$ underlying a second cavity $C_{n+1}$ with $n\in \mathbb{N}^\star$; both are separated by an intermediate barrier. We chose two reference radii $r_n^+$ or $r_{n+1}^-$ that are located below and above the considered barrier, at the interfaces with the resonant cavities $C_n$ and $C_{n+1}$, respectively; these layers define the upper and lower boundaries of the cavities $C_n$ and $C_{n+1}$. Within such a framework, we can first express the wave function inside the cavity $C_n$ in a similar way to \eq{sol general}, but with an explicit representation in terms of the propagation of the energy and with the origin of the wave phase set to the upper boundary, $r_n^+$, that is,
\algn{
\Psi_n(r,t)=a_{n,+}^{(\rightarrow)}~\psi^{(\rightarrow)}(r,t;r_n^+)~+~a_{n,+}^{(\leftarrow)}~\psi^{(\leftarrow)}(r,t;r_n^+)\; ,
\label{Phi sol n}
}
where the superscripts $(\rightarrow)$ and $(\leftarrow)$ denote the wave components in the cavity $C_n$ whose energy propagates upward and downward with the complex amplitudes \smash{$a_{n,+}^{(\rightarrow)}$} and \smash{$a_{n,+}^{(\leftarrow)}$}, respectively. As the phase and group velocities for pressure-dominated (gravity-dominated) modes have the same (opposite) direction, we can directly conclude:%
\algn{
\psi^{(\rightarrow)}(r,t;\bar{r})&=\left \lbrace
\begin{array}{ll}
\psi(r;\bar{r})~e^{-\di \sigma t}   &  \mbox{for pressure modes} \\
\psi^\star(r;\bar{r})~e^{-\di \sigma t} &  \mbox{for gravity modes}
\end{array}
\right. \label{psi energy}\\
\psi^{(\leftarrow)}(r,t;\bar{r})&=\psi^{(\rightarrow)\star}(r,-t;\bar{r}) \label{Phi transf}\; ,
}
where we recall that the $\psi(r;\bar{r})$ function is defined in \eq{psi sol}. The same representation can be used to express the wave function inside the cavity $C_{n+1}$, using the lower boundary $r_{n+1}^-$ as the origin of the wave phase, that is,
\algn{
\Psi_{n+1}(r,t)=a_{n+1,-}^{(\rightarrow)}~\psi^{(\rightarrow)}(r,t;r_{n+1}^-)~+~a_{n+1,-}^{(\leftarrow)}~\psi^{(\leftarrow)}(r,t;r_{n+1}^-) \; ,
\label{Phi sol n+1}
}
where \smash{$a_{n+1,-}^{(\rightarrow)}$} and \smash{$a_{n+1,-}^{(\leftarrow)}$} denote the complex amplitudes of the wave components in the cavity $C_{n+1}$ whose energy propagates upward and downward, respectively. We emphasize that $\Psi_n$ and $\Psi_{n+1}$ are two different representations inside different cavities of the same and unique global wave function $\Psi$, which denotes the solution of the wave equation throughout the star.

\begin{figure}
\centering
\includegraphics[scale=0.25,trim= 0cm 0cm 0cm 0cm, clip]{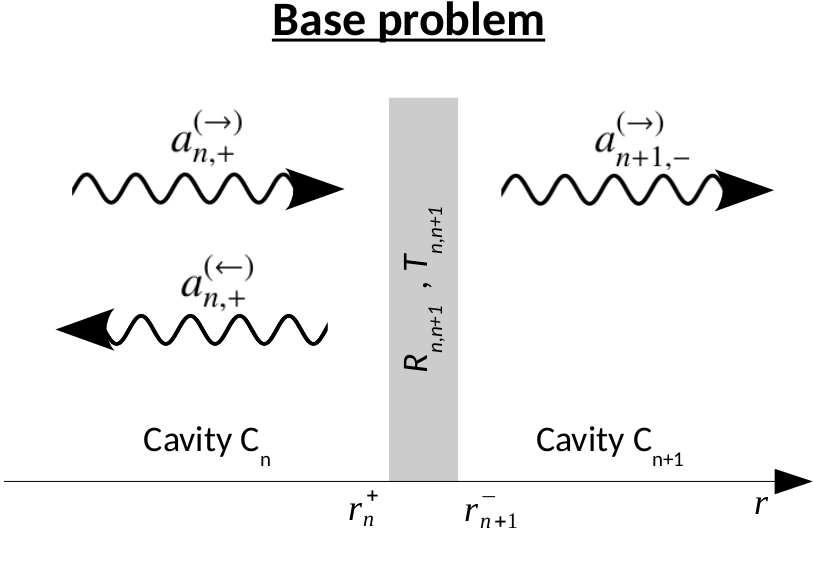} 
\caption{Schematic view of the base wave transmission-reflection problem. An incident wave energy ray in the cavity $C_n$ with an amplitude of \smash{$a_{n,+}^{(\rightarrow)}$} encounters a barrier located between the radii $r_n^+$ and $r_{n+1}^-$, and associated with the (complex) reflection and transmission coefficients $R_{n,n+1}$ and $T_{n,n+1}$. A part of the energy ray is reflected back into the cavity $C_n$ with an amplitude of \smash{$a_{n,+}^{(\leftarrow)}$}, and another part is transmitted into the overlying cavity $C_{n+1}$ with an amplitude of \smash{$a_{n+1,-}^{(\rightarrow)}$}.}
\label{base fig}
\end{figure}

\begin{figure}
\centering
\includegraphics[scale=0.25,trim= 0cm 0cm 0cm 0cm, clip]{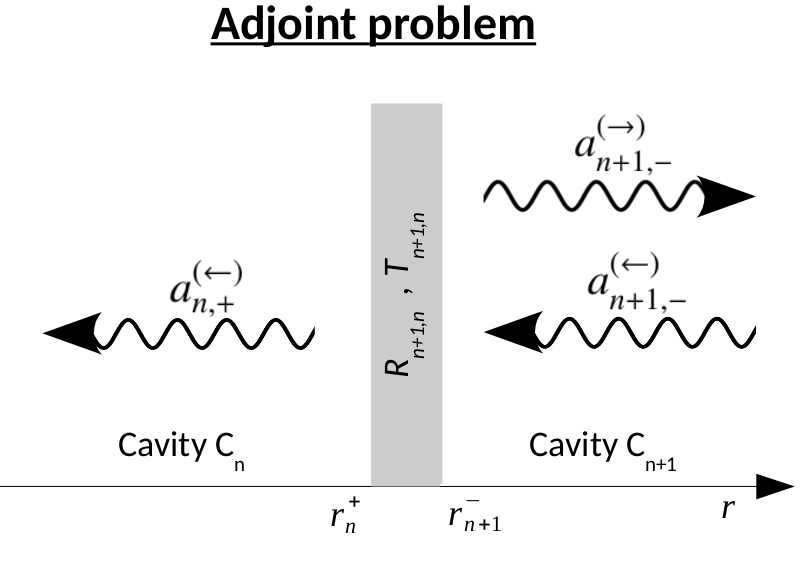} 
\caption{Schematic view of the adjoint wave transmission-reflection problem. An incident wave energy ray in the cavity $C_{n+1}$ with an amplitude of \smash{$a_{n+1,-}^{(\leftarrow)}$} encounters a barrier. A part of the energy ray is reflected back into the cavity $C_{n+1}$ with an amplitude \smash{$a_{n+1,-}^{(\rightarrow)}$}, and another part is transmitted into the underlying cavity $C_{n}$ with an amplitude of \smash{$a_{n,+}^{(\leftarrow)}$}.}
\label{adjoint fig}
\end{figure}

Such a representation of the oscillations in the resonant cavities is appropriate to define properly the reflection and transmission coefficients while accounting for the conservation of the wave energy flux, as already formalized in \cite{Takata2016b}. First, in the base wave reflection-transmission problem, an upward incident energy ray propagating in the bottom cavity $C_n$ encounters the barrier and is reflected back downward, while a part of the energy ray is transmitted through the barrier and propagates upward in the overlying resonant cavity $C_{n+1}$. In other words, we assume \smash{$a_{n+1,-}^{(\leftarrow)}=0$} (see \figurename{}~\ref{base fig}). In this base configuration, the reflection and transmission coefficients, $R_{n,n+1}$ and $T_{n,n+1}$, are respectively defined as the ratio of the amplitudes of the reflected and transmitted components to the amplitude of the incident component inside the resonant cavities, that is,
\algn{
R_{n,n+1}&=\frac{a_{n,+}^{(\leftarrow)}}{a_{n,+}^{(\rightarrow)}}=\left|R_{n,n+1}\right| e^{\di \delta_{n,n+1}} \label{R n n+1}\\
T_{n,n+1}&=\frac{a_{n+1,-}^{(\rightarrow)}}{a_{n,+}^{(\rightarrow)}}=\left|T_{n,n+1}\right| e^{\di \gamma_{n,n+1}} \label{T n n+1}\; ,
}
where $\delta_{n,n+1}$ and $\gamma_{n,n+1}$ are the phase lags introduced at reflection and transmission, respectively. In this work, we will assume that the phase lags take values between $[-\pi,\pi]$. We note that the notion of phase lags is readily defined well inside the cavities where the JWKB is met and where the progressive and regressive wave components are distinctly defined. In addition, the conservation of the mean wave energy flux throughout the star, provided by \eq{mean flux} inside both cavities, translates into the constraint:
\algn{
 \left|a_{n,+}^{(\rightarrow)}\right|^2-\left|a_{n,+}^{(\leftarrow)}\right|^2=\left|a_{n+1,-}^{(\rightarrow)}\right|^2~~ \Rightarrow ~~\left|R_{n,n+1}\right|^2+ \left|T_{n,n+1}\right|^2 =1 \; .
 \label{cons energy}
}
Second, in the adjoint wave reflection-transmission problem, a downward incident energy ray propagating in the resonant cavity $C_{n+1}$ encounters the barrier and is reflected back upward, while a part of the energy ray is transmitted through the barrier and propagates downward in the underlying resonant cavity $C_n$. In other words, we assume \smash{$a_{n,+}^{(\rightarrow)}=0$} (see \figurename{}~\ref{adjoint fig}). In this configuration, the reflection and transmission coefficients, $R_{n+1,n}$ and $T_{n+1,n}$, are  defined, respectively, as
\algn{
R_{n+1,n}&=\frac{a_{n+1,-}^{(\rightarrow)}}{a_{n+1,-}^{(\leftarrow)}}=\left|R_{n+1,n}\right| e^{\di \delta_{n+1,n}}\label{R_n+1_n}\\
T_{n+1,n}&=\frac{a_{n,+}^{(\leftarrow)}}{a_{n+1,-}^{(\leftarrow)}}=\left|T_{n+1,n}\right| e^{\di \gamma_{n+1,n}} \label{T_n+1_n} \; ,
}
where $\delta_{n+1,n}$ and $\gamma_{n+1,n}$ are the phase lags introduced at reflection and transmission. The modulus of the reflection and transmission coefficients must also satisfy the conservation of the wave energy flux as in \eq{cons energy}.

Using basic wave principles of time-reversal symmetry, linear superposition and energy conservation, it is possible using a reasoning similar to \citet[][namely, their Sect.~2,]{Takata2016b} to connect the wave coefficients of the base problem and those of the adjoint problem. As demonstrated in \appendixname{}~\ref{Takata}, they follow the relations
\algn{
\left|R_{n+1,n}\right|&=\left|R_{n,n+1}\right|, &\left|T_{n+1,n}\right|&=\left|T_{n,n+1}\right| \; , \nonumber\\
\gamma_{n+1,n}&=\gamma_{n,n+1},&\delta_{n+1,n}&=\pi-\delta_{n,n+1}+2\gamma_{n,n+1} \label{base adjoint relations}\; ,
}
which is similar to the results found by \cite{Takata2016b} when $\gamma_{n,n+1}=0$.
As a result, we finally see that a barrier between two cavities $C_n$ and $C_{n+1}$ is entirely characterized by only three parameters: $R_{n,n+1}$, $\delta_{n,n+1}$, and $\gamma_{n,n+1}$. The modulus of the transmission coefficient can then be retrieved using the conservation of the wave energy flux.

To complete these definitions, we emphasize that the choice of the origin of the phase in the cavities, that is, the so- called cavity boundaries within our framework, is somehow arbitrary and that any change in the latter is compensated by a modification of the values of the phase lags, $\delta_{n,n+1}$ and $\gamma_{n,n+1}$; the final physical solution of the problem in contrast does not depend on this choice. We note that in practice, it is often convenient to take the boundaries as equal to the turning-points or the middle radius of glitches, as we see later in \sectionname{}~\ref{simple cases}.

 \subsection{Central and surface stellar boundary conditions}
 \label{boundary}
 
Close to the center and surface of stars, the wave energy luminosity must vanish at some point. This condition is required by the regularity of the oscillation displacement near the center and the fact that the density vanishes beyond the surface. The core and surface of stars can thus be modeled as totally reflective barriers \citep[e.g.,][]{Unno1989}.

Inside the first cavity just above the stellar core, which is denoted by $C_1$, the wave function can be generally written within the considered convention as
\algn{
\Psi_{1}(r,t)=a_{1,-}^{(\rightarrow)}~\psi^{(\rightarrow)}(r,t;r_{1}^-)~+~a_{1,-}^{(\leftarrow)}~\psi^{(\leftarrow)}(r,t;r_{1}^-) \; .
\label{Phi sol 1}
}
The reflective boundary condition at the center then requires
 \algn{
 a_{1,-}^{(\rightarrow)} = e^{\di \delta_{\rm c}} a_{1,-}^{(\leftarrow)}\; ,
 \label{core boundary}
 }
where $\delta_{\rm c}$ is the phase lag introduced during the reflection. This ensures that the modulus of the amplitudes of the upward and downward components are equal and hence that the incident wave energy flux is totally reflected.
Similarly, inside the last cavity just below the stellar surface where the stellar density vanishes, which is denoted by $C_N$ with $N\in\mathbb{N}^\star$, the wave function reads
\algn{
\Psi_{N}(r,t)=a_{N,+}^{(\rightarrow)}~\psi^{(\rightarrow)}(r,t;r_{N}^+)~+~a_{N,+}^{(\leftarrow)}~\psi^{(\leftarrow)}(r,t;r_{N}^+) \; .
\label{Phi sol N}
}
The reflective boundary condition at the surface then requires
 \algn{
a_{N,+}^{(\leftarrow)} = e^{\di \delta_{\rm s}} a_{N,+}^{(\rightarrow)}\; ,
\label{upper surface}
 }
 where $\delta_{\rm s}$ is the phase lag introduced at the reflection.
 
At this point, all the basic ingredients have been introduced to formulate in a general way the resonance condition of global modes in stars by taking simultaneously into account the effect of an ensemble of barriers, as we see in the next sections.
%
\section{Infinite-time reflection picture for multi-cavity oscillation modes}
 \label{infinite reflection}

\begin{figure*}
\centering
\includegraphics[width=\textwidth,trim= 0cm 0cm 0cm 1cm, clip]{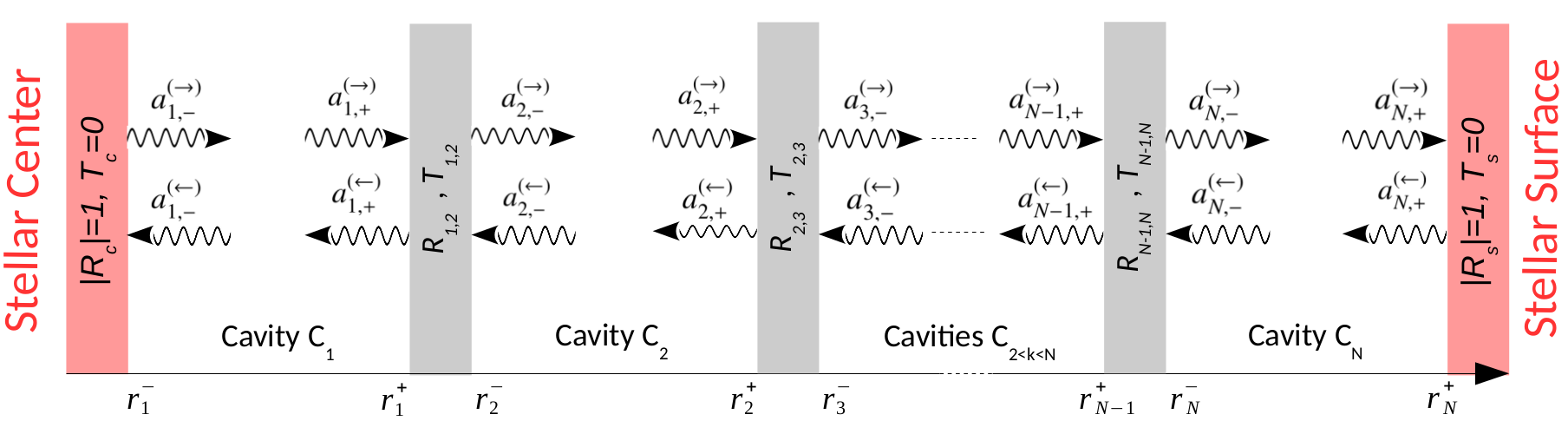} 
\caption{Schematic view of a star modeled as an ensemble of resonant cavities, \smash{$\{C_i\}_{1\le i \le N}$}, where the waves can propagate and the JWKB approximation is met. The cavities are separated from each other by barriers corresponding to either evanescent or rapidly varying regions associated with complex reflection and transmission coefficients $R_{i,i+1}$ and $T_{i,i+1}$ (gray shaded regions). The lower and upper boundaries of the cavity $C_i$ are located at the radii $r_i^-$ and $r_{i}^+$. The upward and downward energy ray in the cavity $C_i$ are associated with the amplitudes \smash{$a_{i,\pm}^{(\rightarrow)}$} and \smash{$a_{i,\pm}^{(\leftarrow)}$}, respectively, where the plus (minus) sign is chosen when the origin of the wave phase in Eqs.~(\ref{phase}) and (\ref{psi sol}) is chosen at $r_i^+$ ($r_{i}^-$). The boundaries at the stellar center and surface are modeled as totally reflective barriers.}
\label{cavity}
\end{figure*}

In this section, we consider a star composed of $N$ successive resonant cavities $\lbrace C_i \rbrace_{1\le i\le N}$ between the center and surface, which are separated from each other by $N-1$ intermediate barriers. In order to make explicit the properties of the oscillation eigenmodes, we follow the propagation of an incident energy ray along its infinite-time travel through the ensemble of cavities and impose a condition for constructive interferences. The problem is graphically represented in \figurename{}~\ref{cavity}. This is actually very similar to the computation of the transmission transfer function of a Fabry-Pérot in optics, but with an ensemble of resonant cavities and totally reflective boundaries.

\subsection{Cavities $C_1$ and $C_2$}
 \label{2 cavities}
 
As a first step, we focus on the two lowest cavities $C_1$ and $C_2$ while considering the reflective condition at the bottom boundary of $C_1$ (i.e., close to the stellar center). In the following, \smash{$\{\tilde{r}_i\}$} denotes a set of given radii in the middle of the cavities $\{C_i\}$. We then consider the case of an incident downward energy ray at \smash{$\tilde{r}_2$} and at a given time $t_0$ with an amplitude \smash{$a_{2,-}^{(\leftarrow)}$}. According to \eq{Phi sol n+1}, the wave function at $\tilde{r}_2$ is provided by
\algn{
\Psi_{2}(\tilde{r}_2,t_0)=a_{2,-}^{(\rightarrow)}~\psi^{(\rightarrow)}(\tilde{r}_2,t_0;r_{2}^-)~+~a_{2,-}^{(\leftarrow)}~\psi^{(\leftarrow)}(\tilde{r}_2,t_0;r_{2}^-) \; .
\label{Phi sol 2}
}
The goal is then to express \smash{$a_{2,-}^{(\rightarrow)}$} as a function of \smash{$a_{2,-}^{(\leftarrow)}$}. Unlike the adjoint wave reflection-transmission configuration presented in \sectionname{}~\ref{barrier}, the amplitude of the reflected component has to account for the reflective boundary condition at the stellar center and the fact that the wave energy that is transmitted from $C_2$ to $C_1$ can come back into $C_2$. This reflected part can thus be decomposed into two components, namely as:
\algn{
a_{2,-}^{(\rightarrow)}=\alpha_{2,R}^{(\rightarrow)}+ \alpha_{2,T}^{(\rightarrow)}\; .
\label{alpha refl}
}
The first term $\alpha_{2,R}^{(\rightarrow)}$ corresponds to the part of the incident energy ray that is directly reflected on the barrier between $C_1$ and $C_2$ as in the associated adjoint wave reflection-transmission problem, that is,
 \algn{
\alpha_{2,R}^{(\rightarrow)}&=R_{2,1}~a_{2,-}^{(\leftarrow)} \label{a_2R}\; .
 }
The second term $\alpha_{2,T}^{(\rightarrow)}$ corresponds to the part of the incident energy ray that is transmitted to $C_1$, that then indefinitely travels back and forth between the center and the upper boundary of $C_1$, and that is finally transmitted back to $C_2$. Based on \eq{Phi sol n}, the wave function at $\tilde{r}_1$ in $C_1$ can be written as
\algn{
\Psi_{1}(\tilde{r}_1,t_0)=a_{1,+}^{(\rightarrow)}~\psi^{(\rightarrow)}(\tilde{r}_1,t_0;r_{1}^+)~+~a_{1,+}^{(\leftarrow)}~\psi^{(\leftarrow)}(\tilde{r}_1,t_0;r_{1}^+) \; .
\label{Phi sol 1}
}
The wave amplitudes in \eq{Phi sol 1} can be expressed as the superposition of an infinite number of contributions resulting from the infinite-time multiple reflections of the energy ray transmitted from $C_2$, that is,
\algn{
a_{1,+}^{(\leftarrow)}&=\alpha_{1,T}^{(\leftarrow)}+\sum_{k=1}^{+\infty} \alpha_{1,k}^{(\leftarrow)} \label{a_1L}\\
a_{1,+}^{(\rightarrow)}&=\alpha_{1,R}^{(\rightarrow)}+\sum_{k=2}^{+\infty} \alpha_{1,k}^{(\rightarrow)}\label{a_1R}\; .
}
First, \smash{$\alpha_{1,T}^{(\leftarrow)}$} corresponds to the part of the incident energy ray that is transmitted from $C_2$ to $C_1$, that is,
\algn{
\alpha_{1,T}^{(\leftarrow)}=T_{2,1}~a_{2,-}^{(\leftarrow)} \label{a_1T}\; .
}
Second, \smash{$\alpha_{1,R}^{(\rightarrow)}$} corresponds to the transmitted part that is in addition reflected close to the core and returns back to $\tilde{r}_1$.
To express this amplitude as a function of the wave coefficients within the conventions presented in \sectionname{}~\ref{setting}, we need to change the origin of the phase in \eq{Phi sol 1} from $r_1^+$ to $r_1^-$ where the reflection occurs. From \eqs{psi sol}{psi energy}{Phi transf}, we deduce for any integer $i$ that
\algn{
\psi^{(\rightarrow)}(\tilde{r}_i,t_0;r_{i}^+)&= e^{-\di \Theta_i}~ \psi^{(\rightarrow)}(\tilde{r}_i,t_0;r_{i}^-) \label{Trans right0},\\
\psi^{(\leftarrow)}(\tilde{r}_i,t_0;r_{i}^+)&=e^{+\di \Theta_i} ~\psi^{(\leftarrow)}(\tilde{r}_i,t_0;r_{i}^-) \label{Trans left0}\; ,
}
with 
\algn{
\Theta_i=\pm \int_{r_{i}^-}^{r_{i}^+} \mathcal{K}_r \dd r \; ,
\label{Theta_1}
}
where the plus (minus) sign has to be chosen when the radial group and phase velocities have the same (opposite) directions, that is, in cases of pressure-dominated (gravity-dominated) modes. In comparing \eq{Phi sol 1} with the expression of the wave function at $\tilde{r}_1$, but with $r_1^-$ as the origin of the wave phase, that is, 
\algn{
\Psi_{1}(\tilde{r}_1,t_0)=a_{1,-}^{(\rightarrow)}~\psi^{(\rightarrow)}(\tilde{r}_1,t_0;r_{1}^-)~+~a_{1,-}^{(\leftarrow)}~\psi^{(\leftarrow)}(\tilde{r}_1,t_0;r_{1}^-) \; ,
\label{Phi sol 1 -}
}
we then can understand, by using \eqsduo{Trans right0}{Trans left0},~that a change in the origin of the phase from the upper boundary of the cavity $C_i$, $r_i^+$, to the lower boundary, $r_i^-$, is associated with the transformation of the amplitudes
\algn{
a_{i,-}^{(\rightarrow)}&=e^{-\di \Theta_i} ~a_{i,+}^{(\rightarrow)}\label{Trans left},\\
 a_{i,-}^{(\leftarrow)}&=e^{+\di \Theta_i} ~a_{i,+}^{(\leftarrow)}\label{Trans right}\; .
} 
Starting from the amplitude \smash{$\alpha_{T,1}^{(\leftarrow)}$}, we first change the origin of the phase to $r_1^-$, which is equivalent to multiply the amplitude by $e^{\di \Theta_1}$ according to \eq{Trans right}. With this convention, the reflected amplitude is just lagged by a phase $\delta_{\rm c}$ compared to the incident one according to \sectionname{}~\ref{boundary}. To conclude, we have to change the origin of the phase back to $r_1^+$ to retrieve the convention used in \eq{Phi sol 1}, which is equivalent to multiply the amplitude again by $e^{\di \Theta_1}$ according to \eq{Trans left}. As a result, this gives
\algn{
\alpha_{1,R}^{(\rightarrow)}=e^{2\di \Theta_1} e^{\di \delta_{\rm c}}\alpha_{1,T}^{(\leftarrow)} \label{a_1RR}\; .
}
Third, $\alpha_{1,k}^{(\leftarrow)}$ and $\alpha_{1,k}^{(\rightarrow)}$ in \eqsduo{a_1L}{a_1R}~correspond to the parts of the downward and upward components having traveled back and forth $k$ times inside $C_1$. For instance, the downward term having made only one back and forth is deduced from the reflection of the amplitude \smash{$\alpha_{1,R}^{(\rightarrow)}$} on the barrier between $C_1$ and $C_2$, that is,
\algn{
\alpha_{1,1}^{(\leftarrow)}&=R_{1,2} \alpha_{1,R}^{(\rightarrow)}=R_{1,2}~e^{2\di \Theta_1} e^{\di \delta_{\rm c}}\alpha_{1,T}^{(\leftarrow)} \; .
\label{alpha_1,1 L}
}
We retrieve a factor $e^{2\di \Theta_1}$ that results from the total "optic" path covered by the initial downward energy ray transmitted from $C_2$ to come back to its initial position during a back and forth in $C_1$. Based on the same reasoning as before, it is straightforward to express the other components by recurrence for $k>1$ as:
\algn{
\alpha_{1,k}^{(\leftarrow)}&=R_{1,2}~e^{2\di \Theta_1} e^{\di \delta_{\rm c}}~\alpha_{1,k-1}^{(\leftarrow)} \label{a_1kL}\\
\alpha_{1,k}^{(\rightarrow)}&=e^{2\di \Theta_1} e^{\di \delta_{\rm c}}~\alpha_{1,k-1}^{(\leftarrow)} \label{a_1kR}\; .
}
As a consequence, the upward and downward amplitudes in $C_1$ can be linked through \eqs{a_1L}{a_1R}{a_1RR}{a_1kR}, which merely results in
\algn{
a_{1,+}^{(\rightarrow)}=e^{2\di \Theta_1} e^{\di \delta_{\rm c}}~a_{1,+}^{(\leftarrow)} \; .
\label{aR_vs_aL}
}
In other words, the upward and downward components in $C_1$ have the same modulus and are just phase lagged during the infinite-time travel of the wave energy ray inside $C_1$. This means that the downward and upward energy rays carry the same amount of energy (but in the opposite direction) in such a way that the total wave luminosity vanishes in $C_1$, as expected from \eq{mean flux} and the conservation of the wave energy flux under the totally reflective core constraint.
Finally, the amplitude of the wave transmitted to $C_2$ is equal to:
\algn{
\alpha_{2,T}^{(\rightarrow)}=T_{1,2}~a_{1,+}^{(\rightarrow)} \; ,
}
so that using \eqss{a_1R}{a_1T}~and \eqss{a_1RR}{a_1kR}, and expressing $T_{2,1}$ as a function of $T_{1,2}$ through \eq{base adjoint relations}, it can be rewritten as:
\algn{
\alpha_{2,T}^{(\rightarrow)}&=a_{2,-}^{(\leftarrow)}~ \left|T_{1,2}\right|^2~e^{\di\left(\delta_{\rm c}+2\gamma_{1,2}+2\Theta_1\right)}\sum_{k=0}^{+\infty} \left[ R_{1,2} e^{\di\left(\delta_{\rm c}+2\Theta_1\right)} \right]^k\nonumber\\
&=a_{2,-}^{(\leftarrow)}~\left|T_{1,2}\right|^2~\frac{e^{\di\left(\delta_{\rm c}+2\gamma_{1,2}+2\Theta_1\right)}}{ 1-R_{1,2} e^{\di \left(\delta_{\rm c}+2\Theta_1\right)}} \; . \label{alpha_2TR}
}
Therefore, the total amplitude of the upward component in $C_2$ is equal to, according to \eqs{base adjoint relations}{alpha refl}{a_2R}{alpha_2TR}:
\algn{
a_{2,-}^{(\rightarrow)}=a_{2,-}^{(\leftarrow)}\left(\left|R_{1,2}\right| e^{\di\left(\pi-\delta_{1,2}+2\gamma_{1,2}\right)}+\frac{\left|T_{1,2}\right|^2~e^{\di\left(\delta_{\rm c}+2\gamma_{1,2}+2\Theta_1\right)}}{ 1-\left|R_{1,2}\right| e^{\di \left(\delta_{\rm c}+\delta_{1,2}+2\Theta_1\right)}}\right) .
\label{eq intermediate}
}
Using the energy constraint $|T_{1,2}|^2=1-|R_{1,2}|^2$, \eq{eq intermediate} leads after some manipulations to
%
%
\algn{
a_{2,-}^{(\rightarrow)}=e^{\di \Delta_{2,1}} a_{2,-}^{(\leftarrow)} \; ,
\label{a_2R=a_2L}
}
where \smash{$\Delta_{2,1}$} represents the total phase lag introduced by the reflection on the intermediate barrier and the infinite-time travel inside $C_1$ of the initial downward incident wave. It reads
 \algn{
 \Delta_{2,1}=\pi+2\gamma_{1,2}-\delta_{1,2}-2 \arctan \left( q_{1,2}\cot \Phi_1\right) \; ,
 \label {Delta 21}
 }
with
 \algn{
  \Phi_1&=\Theta_1+\frac{\delta_{\rm c}}{2}+\frac{\delta_{1,2}}{2} 
 \label{Phi_1}\\
q_{1,2}&=\frac{1-\left|R_{1,2}\right|}{1+\left|R_{1,2}\right|} \; .
 \label{q_12}
 }
The factor $q_{1,2}$ in the last equation is called the coupling factor of the cavities $C_1$ and $C_2$, which can take values between zero and unity.
Equation~(\ref{a_2R=a_2L}) also shows that as in $C_1$, the upward and downward components in $C_2$ have the same modulus because of both the totally reflective core constraint and the energy conservation; the total wave luminosity thus also vanishes in $C_2$.

We thus conclude that the combination of the cavity $C_1$ and the overlying intermediate barrier is equivalent to one single totally reflective barrier underlying the cavity $C_2$ and introducing an effective phase lag $\Delta_{2,1}$ at the reflection of a downward incident wave.

 \subsection{Ensemble of $N$ cavities}
 \label{ensemble}
 
As the next step, we add a third cavity $C_3$ and an intermediate barrier above $C_2$. We then consider the case of an incident downward energy ray at \smash{$\tilde{r}_3$} with an amplitude \smash{$a_{3,-}^{(\leftarrow)}$} at a given time $t_0$. According to \eq{Phi sol n+1}, the wave function at $\tilde{r}_3$ is provided by
\algn{
\Psi_{3}(\tilde{r}_3,t_0)=a_{3,-}^{(\rightarrow)}~\psi^{(\rightarrow)}(\tilde{r}_3,t;r_{3}^-)~+~a_{3,-}^{(\leftarrow)}~\psi^{(\leftarrow)}(\tilde{r}_3,t_0;r_{3}^-) \; .
\label{Phi sol 3}
}
As previously, the goal is to express the total reflected amplitude \smash{$a_{3,-}^{(\rightarrow)}$} as a function of the incident amplitude \smash{$a_{3,-}^{(\leftarrow)}$}. Actually, the computation is similar to that performed in \sectionname{}~\ref{2 cavities} when considering two cavities because we have previously shown that the cavity $C_1$ and the overlying barrier just below $C_2$ together can be represented as a totally reflective barrier associated with an effective phase lag $\Delta_{2,1}$.
Therefore, we can conclude in a straightforward way similarly to \eq{a_2R=a_2L} that
\algn{
a_{3,-}^{(\rightarrow)}=a_{3,-}^{(\leftarrow)} e^{\di \Delta_{3,2}} \; ,
\label{a_3R=a_3L}
}
where $\Delta_{3,2}$ is the total phase lag introduced during the reflection on the barrier between $C_2$ and $C_3$ and the infinite-time travel throughout the cavities $C_1$ and $C_2$. Its expression is provided by the set of \eqss{Delta 21}{q_12}~while replacing $\delta_{\rm c}$ by $\Delta_{2,1}$ and the subscripts $1$ and $2$ by the subscripts $2$ and $3$, respectively (e.g., $|R_{1,2}|$ must be replaced by $|R_{2,3}|$). The region between the stellar core and the barrier underlying the cavity $C_3$ can thus also be considered as a totally reflective barrier associated with an effective phase lag at reflection $\Delta_{3,2}$.

At this point, it is then obvious that the generalization to the case of $N$ cavities can be obtained by adding one by one supplementary overlying cavities and using the same reasoning as before at each step. To do so, we consider a downward incident energy ray in the cavity $C_N$. Analogously to  \eqsduo{a_2R=a_2L}{a_3R=a_3L}, we understand that the amplitudes of the downward and upward components in each cavity are linked for $1\le i \le N$ by the expression
\algn{
a_{i,-}^{(\rightarrow)}=e^{\di \Delta_{i,i-1}}~a_{i,-}^{(\leftarrow)} \; ,
\label{aR_vs_aL i}
}
where the total effective phase lag, $\Delta_{i,i-1}$, introduced by the infinite-time reflections and back-and-forth travels through all the underlying cavities, $\lbrace C_j\rbrace_{j<i}$, can be formulated by the recurrence relation for $i\ge 2:$
 \algn{
 \Delta_{i,i-1}&=\pi+2\gamma_{i-1,i}-\delta_{i-1,i}-2 \arctan \left( q_{i-1,i}\cot \Phi_{i-1}\right) \label{Delta i i-1} \\
   \Phi_{i-1}&=\Theta_{i-1}+\frac{\Delta_{i-1,i-2}}{2}+\frac{\delta_{i-1,i}}{2}
  \label{Phi_i-1} \; .
}
The general definition of the coupling factor between the cavities $C_{i-1}$ and $C_i$ is provided by
\algn{
 q_{i-1,i}&=\frac{1-\left|R_{i-1,i}\right|}{1+\left|R_{i-1,i}\right|} \label{coupling} \; .
}
Finally, the initialization of the recurrence for $i=1$ is ruled by the core boundary condition in \eq{Phi_1}, which is reduced to
\algn{
\Delta_{1,0}=\delta_{\rm c} \; .
\label{initial cond}
}
We also note that \eq{aR_vs_aL i} implies that the modulus of the upward and downward components in all cavities are equal and thus that the total wave energy luminosity is null everywhere. As mentioned before, this is the consequence of both the totally reflective core condition and the mean energy flux conservation.

 \subsection{Upper surface boundary and resonance condition}
 \label{upper resonance}
 
In the final step, we add the surface totally reflective barrier above the cavity $C_N$, so that \eq{upper surface} must apply. Simultaneously, using \eqs{Trans left}{Trans right}{aR_vs_aL i}~for $i=N$, we also have to impose:
\algn{
a_{N,+}^{(\rightarrow)}=e^{2\di \Theta_N} e^{\di \Delta_{N,N-1}}~a_{N,+}^{(\leftarrow)} \; .
}
This provides the following resonance condition:
 \algn{
 \Theta_N+\frac{\delta_{\rm s}}{2}+\frac{\Delta_{N,N-1}}{2} = n \pi \; ,
 \label{resonance}
 }
where $n$ is an integer corresponding to the mode radial order. In the adopted stationary configuration, the resonance condition is actually equivalent to consider that a downward incident energy ray in the cavity $C_N$ has to travel throughout the ensemble of cavities and come back to its initial position with the exact same amplitude in order to constructively interfere. In order to give a usual physical meaning to the radial order $n$ in \eq{resonance}, it may be convenient to choose the branch of the arctangent function in such way that for \smash{any real $\varepsilon$, we have:}
\algn{
\lim_{q\rightarrow1}~ \arctan \left( q \tan\varepsilon\right) = \varepsilon \; .
\label{branch}
}
Using this convention, the radial order $n$ can be interpreted  throughout the paper as the difference between the number of radial oscillation nodes in the p-dominated cavities, $n_{\rm p}$, and that in the g-dominated cavities, $n_{\rm g}$, over all the cavities (i.e., $n=n_{\rm p}-n_{\rm g}$). Using the principal branch for the artangent function, $n$ in \eq{resonance} would be instead interpreted as $n_N$, the number of radial nodes over the cavity $C_N$.

Hence, the mode eigenfrequency spectrum is obtained by solving the set of equations \eqs{Delta i i-1}{Phi_i-1}{initial cond}{resonance}~for any radial order $n$ and accounting for the implicit frequency dependence of the wave number integral and the barrier parameters. To express the resonance condition in a practical form, we can define the phase $\Upsilon_i$ for $1\le i \le N$ such as
\algn{
 \Upsilon_i&=\Theta_i+\frac{\delta_{i,i+1}}{2}+\frac{\delta_{i,i-1}}{2} -\frac{\pi}{2} \; ,
 \label{Ups_i}
 }
and we can set
 \algn{
 \delta_{1,0}=\delta_{\rm c} ~~~~\mbox{and}~~~~\delta_{N,N+1}= \pi+\delta_{\rm s} \; . \label{Ups_1}
}
Using the relation between $\delta_{i,i-1}$ and $\gamma_{i-1,i}$ in \eq{base adjoint relations}, the recurrence resonance relation in \eqsduo{Delta i i-1}{Phi_i-1}~can be written for $2\le i\le N$ in the form of
 \algn{
\tan\Phi_{i-1}=q_{i-1,i}  \tan \left( \Phi_{i}-\Upsilon_i\right)\; ,
 \label{tan Phi}
 }
where the core and surface boundary conditions in \eqsduo{initial cond}{resonance}~translate respectively into
 \algn{
\Phi_1=\Upsilon_1+\frac{\pi}{2}~~~~\mbox{and}~~~~ \Phi_N= \left(n+\frac{1}{2}\right)\pi\; . \label{cond tan Phi}
 }
For $N=1$, \eq{cond tan Phi} is sufficient alone and leads to \smash{$\Upsilon_1=n\pi$.}
Depending on the problem, it can also be convenient to express the resonance condition as a series of frequency-dependent sine terms, as we see in \sectionname{}~\ref{glitch}. Such an alternative formulation is provided in \appendixname{}~\ref{sine}.

 \subsection{On the need for the multi-cavity approach}

Before going further, it is worth discussing the need for the multi-cavity approach, since we show in Sects.~\ref{2 cavities} and~\ref{ensemble} that it is possible to reinterpret the multi-cavity problem as the single-cavity problem. Indeed, the combination of the cavities and barriers below a given cavity $C_{i>1}$ can always be reinterpreted as one single totally reflective barrier with an effective reflection phase lag redefined appropriately. Actually, the same conclusion holds true for the combination of the cavities and barriers above a given cavity, $C_{i<N}$. This can be easily shown following the same reasoning as in Sects.~\ref{2 cavities} and~\ref{ensemble}, except that the computation of the mode amplitude has to be made starting from the surface totally reflective condition and going toward deeper cavities. We may therefore wonder whether the multi-cavity picture, which is more complicated, is justified or not. The answer to this question depends on how the wave number integrals and the reflection coefficients vary with frequency over a considered range. On the one hand, if the wave number integral and the reflection coefficients on both sides of the considered cavity are constant over the frequency range of interest, then the single-cavity picture appears sufficient to describe the oscillations. On the other hand, if the mode parameters vary even slowly with frequency, then we can expect that the effective phase lags resulting from the combination of the cavities above and below the considered cavity behave in a complicated way with frequency. In this case, the use of the single-cavity picture is not judicious. In practice, for real stars, the wave number integrals and the reflection coefficients always vary with frequency and choosing a multi-cavity approach to describe the oscillation modes therefore appears necessary to understand their frequency spectra and develop useful seismic diagnoses.

%
\section{Linear boundary value problem picture}
 \label{BV picture}
 
In \sectionname{}~\ref{infinite reflection}, we describe the eigenmodes using a physical picture that is analogous to ray tracing in optics. In this section, we aim to describe the eigenmodes in a more mathematical way that considers them as the solution to a linear boundary value problem. In addition, to check the validity of the previous scenario, such an approach also has the advantage of being more convenient to discuss the distribution of the mode energy and the mode amplitudes ratios throughout stars, which is addressed in \sectionname{}~\ref{mode energy and amplitude}.

 \subsection{Amplitude vector}
 
 In each resonant cavity, the general solution for the wave function $\Psi$ takes a similar form to \eqsduo{Phi sol n}{Phi sol n+1}~depending on whether the origin of the phase is taken at the lower or upper boundary of the cavity, respectively. In each cavity $C_i$, we thus define the amplitude vector in both cases as
 \algn{
 \vec{a}_{i,-}= \left(
\begin{array}{c}
a_{i,-}^{(\rightarrow)}\\
a_{i,-}^{(\leftarrow)}
\end{array}
\right)~~~~\mbox{and}~~~~
\vec{a}_{i,+}= \left(
\begin{array}{c}
a_{i,+}^{(\rightarrow)}\\
a_{i,+}^{(\leftarrow)}
\end{array} \right)\; .
 }
To find the stationary modes oscillating between the central and surface boundaries of the star, we need to make the link between all the amplitude vectors from the cavities $C_1$ to $C_N$.
 
 \subsection{Connection through the intermediate barriers}
 
 Around each barrier located between the cavities $C_i$ and $C_{i+1}$, a first solution for the wave function is the solution of the base wave reflection-transmission problem presented in \sectionname{}~\ref{barrier} and is denoted by $\Psi^{(1)}$. This solution is associated in both cavities with the amplitude vectors up to a given proportionality constant:
 \algn{
 \vec{a}_{i+1,-}^{(1)}\propto \left(
\begin{array}{c}
T_{i,i+1}\\
0
\end{array}
\right)~~~~\mbox{and}~~~~
\vec{a}_{i,+}^{(1)}\propto \left(
\begin{array}{c}
1\\
R_{i,i+1}
\end{array} \right)\; .
\label{eq a_-}
 }
As shown in \appendixname{}~\ref{Takata}, another solution can then be obtained by complex conjugation and time reversal. In other words, $\Psi^{(2)}=\Psi^{(1)\star}(r,-t)$ is also a solution. According to \eq{Phi transf} (see also the example in \appendixname{}~\ref{Takata}), this solution is associated with the amplitude vectors, up to the same proportionality constant as in \eq{eq a_-},
 \algn{
 \vec{a}_{i+1,-}^{(2)}\propto \left(
\begin{array}{c}
0\\
T_{i,i+1}^\star
\end{array}
\right)~~~~\mbox{and}~~~~
\vec{a}_{i,+}^{(2)}\propto \left(
\begin{array}{c}
R_{i,i+1}^\star\\
1
\end{array} \right)\; .
 }
The two solutions $\Psi^{(1)}$ and $\Psi^{(2)}$ are linearly independent and constitute a basis for the solution of the wave equation around the considered barrier (indeed, it is straightforward to show that their Wronskian is not null). They are thus sufficient to deduce the general transformation making the connection between $\vec{a}_{i+1,-}$ and $\vec{a}_{i,+}$ for $1\le i \le N-1$ in the following matrix form
\algn{
\vec{a}_{i,+}=\mathsf{B}_{i,i+1}~\vec{a}_{i+1,-} \; ,
\label{connection}
}
where $\mathsf{B}_{i,i+1}$ is the transformation matrix:
\begingroup\addtolength{\jot}{0.2cm}
\algn{
\mathsf{B}_{i,i+1}&= \left(
\begin{array}{cc}
T_{i,i+1}^{-1}&R_{i,i+1}^\star T_{i,i+1}^{\star ~-1}\\
R_{i,i+1} T_{i,i+1}^{ -1}&T_{i,i+1}^{\star ~-1}
\end{array} \right)\nonumber\\
&= \mathsf{C}\left(-\frac{\delta_{i,i+1}}{2} \right)  \mathsf{A}_{i,i+1} \mathsf{C}\left(\frac{\pi}{2} -\frac{\delta_{i+1,i}}{2}\right) \; ,
\label{B matrix}
}
where we have used \eq{base adjoint relations}  to decompose the matrix and where we have defined
\algn{
\mathsf{C}(\varepsilon)&= \left(
\begin{array}{cc}
e^{+\di \varepsilon}&0\\
0&e^{-\di \varepsilon}
\end{array} \right) \label{C matrix}\\
\mathsf{A}_{i,i+1}&= \frac{1}{\left|T_{i,i+1}\right|}\left(
\begin{array}{cc}
1&\left|R_{i,i+1}\right|\\
\left|R_{i,i+1}\right|&1
\end{array} \right) \; . \label{A matrix}
}
\endgroup
For our purpose, it is useful to specify how the $\mathsf{C}$ and $\mathsf{A}_{i,i+1}$ matrices transform a given amplitude vector in the form of:
\algn{
\vec{\varw}(\varepsilon) =\left(
\begin{array}{c}
e^{+\di \varepsilon}\\
e^{-\di \varepsilon}
\end{array} \right) \; . \label{w vector}
}
It is straightforward to find~that
\algn{
&\mathsf{C}(\varepsilon_2)~\vec{\varw}(\varepsilon_1)=\vec{\varw}(\varepsilon_2+\varepsilon_1) \label{C transf}\\
&\mathsf{A}_{i,i+1}~\vec{\varw}(\varepsilon) = \mathcal{A}_{i,i+1} (\varepsilon)~ \vec{\varw}\left( \arctan\left[q_{i,i+1} \tan\left(\varepsilon \right) \right]\right) \; ,
\label{A transf}
}
where, using the convention in \eq{branch} for the branch of the arctangent function\footnote{When using the principal branch of the actangent function, \eq{A amplitude} has to be multiplied by an additional factor of ${\rm sgn}(\cos(\varepsilon))$.},
\algn{
\mathcal{A}_{i,i+1} (\varepsilon)
&= \sqrt{\frac{q_{i,i+1}^2+(1-q_{i,i+1}^2)\cos^2 \left( \varepsilon \right)}{q_{i,i+1}}}\; .
\label{A amplitude}
}
%

\subsection{Wave function matching in the middle of the cavities}
 
At this point, we need to connect the general solution around the barrier between $C_{i}$ and $C_{i+1}$ to the general solution around an adjacent barrier, for instance, between $C_{i-1}$ and $C_i$. It is thus sufficient to make the link between $\vec{a}_{i,+}$ and $\vec{a}_{i,-}$ inside the cavity $C_i$ for $1\le i \le N$. This is provided by \eqsduo{Trans left}{Trans right}, and can be written in the following matrix form:
 \algn{
 \vec{a}_{i,-}= \mathsf{C}_i ~\vec{a}_{i,+} \equiv\mathsf{C}(-\Theta_i)~\vec{a}_{i,+}\; , \label{matching}
 }
where $\Theta_i$ and $\mathsf{C}$ are defined in \eqsduo{Theta_1}{C matrix}. This simple transformation is actually equivalent to match the two JWKB solutions for the wave functions coming from $r_i^-$ and $r_i^+$, respectively, inside the cavity $C_i$, as usually done for instance in the usual asymptotic analyses of stellar pulsations \citep[e.g.,][]{Shibahashi1979,Tassoul1980,Takata2016a}.
 
\subsection{General transformation between adjacent cavities}

The two last operations can be composed to describe the general transformation of the amplitude vector from a cavity to an adjacent cavity, that is, the transformation from $\vec{a}_{i+1,+}$ to $\vec{a}_{i,+}$. 
Using \eqsduo{connection}{matching}, we get for $1\le i \le N-1$ :
\algn{
 \vec{a}_{i,+}= \mathsf{B}_{i,i+1}\mathsf{C}_{i+1} ~\vec{a}_{i+1,+} \equiv \mathsf{E}_{i,i+1}~ \vec{a}_{i+1,+}.
 \label{general transf}
}
It is also useful for the following to deduce the transformation of a vector amplitude $\vec{\varw}(\varepsilon)$ as defined in \eq{w vector} by the matrix $\mathsf{E}_{i,i+1}$. Using \eqs{B matrix}{C transf}{A transf}{matching}, we find for \smash{$1\le i \le N-1$} that
\algn{
\mathsf{E}_{i,i+1}~ \vec{\varw}(\varepsilon)= \mathcal{A}_{i,i+1} \left(\varepsilon-\Upsilon_{i+1}+\frac{\delta_{i+1,i+2}}{2}\right) ~\vec{\varw}\left[ \varOmega_{i+1}(\varepsilon)\right] \; ,
\label{E w}
}
with
\algn{
\varOmega_{i+1}(\varepsilon)=\arctan\left[q_{i,i+1} \tan\left(\varepsilon-\Upsilon_{i+1}+\frac{\delta_{i+1,i+2}}{2} \right) \right] -\frac{\delta_{i,i+1}}{2} \; ,
\label{Omega}
}
where we have used \eqsduo{base adjoint relations}{Ups_i}~to express the result as a function of the phase $\Upsilon_i$.

\subsection{Surface boundary condition}
\label{surface boundary}

We first apply the totally reflective boundary condition at the surface of the star, which is represented by \eq{upper surface}. In terms of the amplitude vector in $C_N$, it reads:
\algn{
\vec{a}_{N,+} = a_N ~\vec{\varw}\left(-\frac{\delta_{\rm s}}{2} \right)\; ,
\label{surface vector}
 }
where $a_N$ is a complex constant. Imposing this surface condition, it is then possible to deduce the amplitude vectors in the underlying cavities by successively applying the linear general transformation $E_{i,i+1}$ in \eq{general transf} from $i=N-1$ to $i=k$. This gives
\algn{
\vec{a}_{k,+}&=\left(\prod_{i=k}^{N-1} \mathsf{E}_{i,i+1} \right)a_N~ \vec{\varw}\left(-\frac{\delta_{\rm s}}{2} \right) \label{a_k prod} \; .
}
Using \eq{E w} and setting for convenience
\algn{
\varOmega_{N+1}(\varepsilon)=\varepsilon-\frac{\delta_{\rm s}}{2}\; ,
}
this can be expressed in the simple form
\algn{
\vec{a}_{k,+}=a_k ~ \vec{\varw} \left[  \varOmega_{k+1} \circ \cdot \cdot \cdot \circ \varOmega_{N+1} (0) \right] \; , \label{a_k prod 2}
}
where $(\circ)$ is the composition operator and $a_k$ is given by
\algn{
\frac{a_k}{a_N}=\prod_{i=k}^{N-1}  \mathcal{A}_{i,i+1} \left[\varOmega_{i+2} \circ \cdot \cdot \cdot \circ \varOmega_{N+1} (0)-\Upsilon_{i+1}+\frac{\delta_{i+1,i+2}}{2}\right] \; ,
\label{a_k/a_N}
}
We note, according to \eq{a_k prod 2}, that in a given cavity, the amplitudes of the upward and downward components have the same modulus. This is the result of the totally reflective surface boundary condition in \eq{surface vector} and has already been demonstrated using the infinite-time reflection picture.
\subsection{Central boundary and eigenvalue conditions}

In the final step, we apply the bottom boundary condition close to the core of the star, which is represented by \eq{core boundary}. In terms of the vector amplitude, this translates into
\algn{
\vec{a}_{1,-} = \tilde{a}_1 ~ \vec{\varw}\left(\frac{\delta_{\rm c}}{2} \right)\; , \label{a_1 boundary}
 }
where $\tilde{a}_1$ is  a complex constant. Another expression of $\vec{a}_{1,-}$ can also be obtained but satisfying, this time, the surface boundary condition. Using \eqsduo{matching}{a_k prod 2}, it is provided by
\algn{
\vec{a}_{1,-}=\mathsf{C}_1~\vec{a}_{1,+}=a_1 ~ \vec{\varw} \left[  \varOmega_{2} \circ \cdot \cdot \cdot \circ \varOmega_{N+1} \left(0\right)-\Theta_1 \right] \; .\label{a_1 matching}
}
Equations~(\ref{a_1 boundary}) and (\ref{a_1 matching}) have therefore to be met simultaneously to obtain a solution of the boundary value problem. The eigenvalue condition is thus met if the amplitudes are chosen such as \smash{$\tilde{a}_1 = (-1)^{n} a_1$} and 
\algn{
\varOmega_{1} \circ \cdot \cdot \cdot \circ \varOmega_{N+1} (0) = n\pi\; ,
\label{resonance math}
}
where we have defined
\algn{
\varOmega_{1}(\varepsilon)=-\varepsilon+\Upsilon_{1}-\frac{\delta_{1,2}}{2} +\frac{\pi}{2} \; ,
\label{Omega_1}
}
and where $n$ is the radial order.
It is straightforward to show that \eq{resonance math} is equivalent to the recurrence condition found in the infinite-time reflection picture and provided by \eqsduo{tan Phi}{cond tan Phi}. Both approaches are therefore equivalent. The boundary value problem scenario nevertheless turns out to be more convenient to discuss the mode amplitude, as we show in the next section.

%
\section{Local mean mode energy and amplitudes}
 \label{mode energy and amplitude}
 
In addition to the eigenfrequencies, it is also interesting to know the distribution of the mode energy throughout the ensemble of cavities. This can provide information about the regions that the modes can efficiently probe and this is essential for predicting the surface mode displacement and, thus, the observed oscillation power spectra \citep[e.g.,][]{Chaplin2005}. In this section, we briefly address this point based on the present basic formulation of eigenmodes.
 
 \subsection{Mean mode energy in each cavity}
 
Owing to the equipartition of the potential and kinetic energy of short-wavelength gravito-acoustic waves \citep [e.g.,][]{Lighthill1978}, the mean mode energy averaged over one oscillation period \smash{$T=2\pi/\sigma$} in the cavity $C_k$ is defined as:
 \algn{
\mathcal{E}_k = \frac{1}{T} \int_{-T/2}^{+T/2} \left( \int_{r_k^-}^{r_k^+} \iint_{\Sigma} \rho \left| \mathcal{R}_{\rm e} \left[ \di \sigma\vec{\xi}(\vec{r},t)\right] \right|^2 r^2 \dd^2 \Sigma \dd r \right) \dd t \; ,
\label{local energy}
 }
where $\dd^2\Sigma=\sin \theta \dd \theta \dd s$ is the solid angle in the direction $(\theta,s)$. First, to express \eq{local energy}, we note that the horizontal mode displacement can be related to the Eulerian pressure perturbation through the momentum conservation in the horizontal direction and the equality \citep[e.g.,][]{Unno1989}:
 \algn{
 \tilde{\xi}_h=\frac{\tilde{p}^\prime}{\rho r \sigma ^2} \; .
 \label{xi_h}
 }
Then, using \eqs{xi pert}{v_r xi}{v_r}{p^prime}{xi_h}, and taking advantage of the properties of the spherical harmonics,
we find at leading order for \smash{$\sigma^2 \gg (S_\ell^2~\mbox{and}~N^2)$} or \smash{$\sigma^2 \ll (S_\ell^2~\mbox{and}~N^2)$} that (see \appendixname{}~\ref{energy} for details)
\algn{
\mathcal{E}_k \approx |a_k|^2 |\Theta_k| \; .
\label{energy approx}
}
The local mean mode energy is therefore first directly proportional to the wave number integral in the cavity $C_k$ defined in \eq{Theta_1}; we can approximately write $\Theta_k \approx \pi n_k$, where $n_k$ is an integer whose modulus represents the number of oscillation nodes in the radial direction in the cavity $C_k$. Second, it is proportional to the squared modulus of the mode amplitude. Knowing the actual value of this amplitude would require studies of the excitation and damping of the modes, which is beyond the scope of this work. The previous analysis can nevertheless provide us with the amplitude ratios between adjacent cavities, as we show in the next section.
 
 \subsection{Amplitude ratios}
 \label{ratio}
 
Using \eqsduo{A amplitude}{a_k/a_N}, the squared amplitude in the cavity $C_k$ relatively to that in the surface cavity $C_N$ is equal to  
 \algn{
 \left|\frac{a_k}{a_N}\right|^2 = \prod_{i=k}^{N-1} \frac{q_{i,i+1}^2+(1-q_{i,i+1}^2) \cos^2\left( \Xi_{i+1} \right)}{q_{i,i+1}} \; ,
 }
 where we define the quantity
 \algn{
 \Xi_{i+1}=\varOmega_{i+2} \circ \cdot \cdot \cdot \circ \varOmega_{N+1} (0)-\Upsilon_{i+1}+\frac{\delta_{i+1,i+2}}{2} \; .
 }
 The squared amplitude ratio between the adjacent cavities $C_k$ and $C_{k+1}$ is thus merely equal to
 \algn{
 \left|\frac{a_k}{a_{k+1}}\right|^2= \frac{q_{k,k+1}^2+(1-q_{k,k+1}^2) \cos^2\left( \Xi_{k+1} \right)}{q_{k,k+1}} \; . \label{amp ratio}
 }
We can thus see that the squared amplitude ratio of the wave function between adjacent cavities depends on two ingredients: first, the coupling factor associated with the intermediate barrier, which measures how the energy is transmitted from one cavity to the other cavity; second, the ``optical'' path during a back and forth travel of a wave inside the ensemble of cavities above the intermediate barrier, which is represented by $\Xi_{k+1}$ and measures the level of constructive interference in the cavity $C_{k+1}$. In the case when $q_{k,k+1}$ is close to unity, \eq{amp ratio}~show that $|a_k/a_{k+1}|$ is on the order of unity. Indeed, in this case, both cavities are well coupled and exchange comparable wave energy fluxes. In contrast, when $q_{k,k+1}$ is much smaller than unity, the result depends on the level of constructive interference in the cavity $C_{k+1}$. In the limiting case where the level of constructive interference is maximum, which is equivalent to $\Xi_{k+1} = (n_{k+1}+1/2) \pi$ with $n_{k+1}$ a given integer, \eq{amp ratio} shows that the squared amplitude of the wave function is smaller in the cavity $C_k$ than in the cavity $C_{k+1}$ by a factor of $q_{k,k+1}\ll 1$. Conversely, when the level of constructive interference is minimum, that is $\Xi_{k+1} = n_{k+1} \pi$, the squared amplitude of the wave function is larger in the cavity $C_k$ than in the cavity $C_{k+1}$ by a factor of $1/q_{k,k+1}\gg 1$. 

Finally, we first recall that the previous discussion address the amplitude ratio of the wave function in both cavities and that, in order to translate these results in terms of mode energy, we also have to take the $\Theta_k$ factor in \eq{energy approx} into account, which represents the local mode inertia in the asymptotic limit. Second, it is worth mentioning that \eq{amp ratio} expresses $a_k$ as a function of the mode amplitude in the upper cavity $C_N$ since it was obtained by imposing only the surface boundary condition; it is thus valid under this sole condition, which is met by definition for eigenmodes (i.e., in the resonance condition). Imposing instead the sole core boundary condition, it is also possible to derive an expression of $a_k$ as a function of  the mode amplitude $a_1$ in the inner cavity $C_1$, as shown in \appendixname{}~\ref{local-to-core}. Such an expression is equivalent to \eq{amp ratio} only and only for eigenmodes.

%
\section{Simple cases}
\label{simple cases}

In this section, we apply the present formulation on simple cases and check the compatibility with the eigenfrequency conditions already obtained in previous works.

\subsection{Gravity and acoustic modes with one single cavity}
\label{pure g and p}

First, we consider the case of pure gravity or acoustic modes propagating in one single cavity. The cavity is supposed to be located between two turning-points close to the core and surface and beyond which the modes are evanescent, in agreement with the totally reflective boundary conditions. Using \eq{cond tan Phi}, or equivalently \eq{resonance math} for $N=1$, we generally find that:%
\algn{
\int_{r_1^-}^{r_1^+} \mathcal{K}_r\dd r\pm\left(\frac{\delta_{\rm c}}{2}+\frac{\delta_{\rm s}}{2}\right)=\pm n\pi \; ,
\label{1 cavity}
}
where the plus and minus signs correspond to the case of acoustic and gravity modes, respectively, and $n$ is a positive integer representing the mode radial order. 
In general, the values of the phase lags $\delta_{\rm s}$ and $\delta_{\rm c}$ can be computed in a second step using a single turning point asymptotic analysis, such as that performed by \cite{Shibahashi1979}. Retaining the Cowling approximation and neglecting the gradients in the equilibrium structure in the wave equation everywhere inside the star, \cite{Shibahashi1979} showed that the wave function, $\Psi,$ takes the form of an Airy function of the first kind in the vicinity of a given single turning point. Choosing the origins of the wave phase in \eq{phase} as equal to the turning points, that is, such as $\mathcal{K}_r(r_1^-)=\mathcal{K}_r(r_1^+) =0$, we show in \appendixname{}~\ref{phase lag} that $\delta_{\rm c}+\delta_{\rm s}=\pi$ if the core and surface turning points have the same nature (i.e., both turning points satisfy either $\sigma^2=N^2$ or $\sigma^2=S_\ell^2$); and then $\delta_{\rm c}+\delta_{\rm s}=0$ otherwise.

As an illustration, we can consider the examples of the low-frequency gravity modes and high-frequency acoustic modes in low-mass main sequence stars. For the former, the core and surface turning points (i.e., $r_1^-$ and $r_1^+$ in the chosen convention) are such that $\sigma^2=N^2(r_1^-)=N^2(r_1^+)$ \citep[e.g.,][]{Appourchaux2010}. In this case, we thus have:
\algn{
\delta_{\rm c}+\delta_{\rm s}=\pi \; .
}
For the latter, the core and surface turning points are such that $\sigma^2=S_\ell^2(r_1^-)$ and $\sigma^2=N^2(r_1^+)$, so that we get:
\algn{
\delta_{\rm c}+\delta_{\rm s}=0 \; .
\label{p resonance}
}
We note that we retrieve the same result as \cite{Shibahashi1979} for high-frequency acoustic modes, but that we find an additional phase lag of $\pi/2$ in the quantization condition of low-frequency gravity modes. This results from the fact that \cite{Shibahashi1979} considered in contrast that the upper turning point is such that $\sigma^2=S_\ell^2(r_1^+)$. 

When accounting for the perturbation of the gravitational potential, that is in the non-Cowling case, \cite{Takata2005,Takata2006a} demonstrated for the $\ell=1$ modes that the nature of the core turning point changes. For high-frequency acoustic modes (respectively, low-frequency gravity modes), the situation is actually similar to assume simultaneously the Cowling approximation and $\sigma^2=N^2(r_1^-)$ (respectively, $\sigma^2=S_\ell^2(r_1^-)$); as a consequence, $\delta_{\rm c}+\delta_{\rm s}=0$ (respectively, $\delta_{\rm c}+\delta_{\rm s}=\pi$), that is, with a phase shift of $\pi$ compared to within the Cowling approximation \citep{Takata2016a,Pincon2019}. This point emphasizes the importance to take the perturbation of the gravitational potential into account for dipolar modes to study the phase lag at reflection and transmission inside the deep interior of stars.

\subsection{Two-cavity mixed modes}
\label{MM}

In a second example, we consider the case of mixed modes, which can exist not only in red giant stars where they have already been detected, but also in main sequence stars or the Sun. Mixed modes can propagate through an inner buoyancy cavity $C_1$, where they behave as gravity modes, and an outer pressure cavity $C_2$, where they behave as acoustic modes. Using \eqsduo{tan Phi}{cond tan Phi}, or, equivalently, \eq{resonance math} for $N=2$, we can easily get the general resonance condition that states:
\algn{
\cot\left(-\Theta_1-\frac{\delta_{\rm c}}{2}-\frac{\delta_{1,2}}{2} +\frac{\pi}{2}\right)\tan\left(\Theta_2+\frac{\delta_{\rm s}}{2}+\frac{\delta_{2,1}}{2} \right)=q_{1,2} \; .
\label{resonance MM}
}
Equation~(\ref{resonance MM}) can be shown to be similar to Eq.~(18) of \cite{Takata2016b} using the relation $\delta_{2,1}=\pi-\delta_{1,2}+2\gamma_{1,2}$, with $\gamma_{1,2}=0$. 

At this point, we can also discuss the value of the phase lags in the case of an evolved red giant star within the Cowling approximation, following again the work by \cite{Shibahashi1979}. In such considerations, the cavity $C_1$ is located between two turning points such as $\sigma^2=N^2(r_1^-)=N^2(r_1^+)$, and the cavity $C_2$ is located between a lower and upper turning points such as $\sigma^2=S_\ell^2(r_2^-)$ and $\sigma^2=N^2(r_2^+)$, respectively \citep[e.g.,][]{Hekker2017}. The boundaries of the cavities are chosen equal to the turning points. The main assumption of the analysis of \cite{Shibahashi1979} then consists in considering that the four turning points are far away from each other; this permits us to exploit the results obtained from a single turning point analysis of the stellar oscillation equations around each of them. Within this context, neglecting the gradients of the equilibrium structure in the wave equation everywhere, we can deduce $\delta_{\rm s}=-\delta_{\rm c}=-\pi/2$ according to \appendixname{}~\ref{phase lag} and \eq{delta_t 2}. Moreover, as the intermediate evanescent region is thick (i.e., weak coupling between $C_1$ and $C_2$), the phase lags associated with the intermediate evanescent barrier $\delta_{1,2}$ and $\delta_{2,1}$ can also be shown to follow the same rules as in \appendixname{}~\ref{phase lag} \citep[see, e.g.,][]{Takata2016a,Pincon2019}. We therefore deduce $\delta_{1,2} \approx\delta_{2,1}\approx\pi/2$ and $\gamma_{1,2}=0$. In other words, in the Cowling and weak coupling paradigm, \eq{resonance MM} is reduced to $\cot (-\Theta_1)\tan \Theta_2=q_{1,2}$, which is similar to Eq.~(31) of \cite{Shibahashi1979}, since we recall that $\Theta_1$ is minus the wave number integral in the cavity $C_1$. To be precise, we also first note that in the non-Cowling case, we must apply a shift of $\pi$ on $\delta_{\rm c}$, so that $\delta_{\rm c}=-\pi/2$. Moreover, the more complex expressions of $\delta_{1,2}$ and $\delta_{2,1}$ in the approximation of a very thin evanescent region (i.e., strong coupling hypothesis) are also available \citep{Takata2016a}.
 
\subsection{Three-cavity mixed modes}

More recently, \cite{Cunha2015,Deheuvels2018} tackled the case of mixed modes propagating in three resonant cavities. Using \eqsduo{tan Phi}{cond tan Phi}, or equivalently \eq{resonance math}~for $N=3$, we find for three cavities $C_1$, $C_2$, and $C_3$ that the resonance condition reads in a general way
\algn{
\tan &\Upsilon_3 \left(1-q_{1,2} \tan \Upsilon_2 \tan \Upsilon_1\right) +q_{2,3}\tan \Upsilon_2 +q_{1,2}q_{2,3}\tan \Upsilon_1 =0 \; .
\label{resonance 3M}
}
When $q_{2,3}=0$, we note that the term in the brackets is equal to zero and we retrieve the resonance condition in \eq{resonance MM} for two-cavity eigenmodes with the substitution $\delta_{2,3}\leftarrow \delta_{\rm s}$.

As an additional check, it is interesting to compare in more details this result with the asymptotic analysis of mixed modes by \cite{Deheuvels2018} in the case of helium-core flash red giant stars. In contrast with less evolved red giant stars, the presence of a convective region at the border of the helium core locally creates an evanescent region that splits the radiative core into two cavities. where the modes behave as gravity modes \citep[see, e.g., Fig.~2 of][]{Deheuvels2018}. Assuming the Cowling approximation, the turning points of the two inner cavities are such as $\sigma^2=N^2$ while the configuration in the upper cavity is similar to the case of acoustic modes studied in \sectionname{}~\ref{MM}. Choosing the turning points as the origin of the wave phase and using the low-coupling hypothesis in both evanescent regions, the same arguments as in \sectionname{}~\ref{MM} thus hold true and we can write
\algn{
\delta_{\rm c}+\delta_{1,2}=\delta_{2,1}+\delta_{2,3}=\pi~~~~\mbox{and}~~~~\delta_{3,2}+\delta_{\rm s}=0 \; .
\label{delta MM}
}
Therefore, using \eqsduo{Ups_i}{Ups_1}~with \eq{delta MM}, \eq{resonance 3M} can be expressed as
\algn{
\cot & \left( -\Theta_1\right)\cot  \left( -\Theta_2\right)\tan \Theta_3 - q_{1,2} \tan \Theta_3 -q_{2,3} \cot\left( -\Theta_1\right)\nonumber\\
&-q_{1,2}q_{2,3} \cot\left( -\Theta_2\right)=0 \; .
\label{resonance 3M f}
}
We retrieve here, based on basic arguments, the same expression as in Eqs.~(9) and (10) of \cite{Deheuvels2018}, keeping in mind the fact that the minus sign in front of $\Theta_1$ and $\Theta_2$ comes from the definition of $\Theta_i$ in \eq{Theta_1} for gravity-dominated modes. We finally note that \cite{Cunha2015,Cunha2019} formulated the resonance condition in an equivalent way to \eq{resonance 3M}, but, in addition, these authors expressed the $q_{1,2}$ factor by assuming that the barrier corresponds to either a sharp Dirac or Gaussian peak in the Brunt-Väisälä frequency.

\subsection{Low-amplitude glitches}
\label{glitch}

As a final illustration, we consider the important case of low-amplitude glitches, that is, low-amplitude sharp and very localized features in the equilibrium structure. In most previous studies, the effect of such sharp gradients on the eigenfrequencies was usually treated as a small perturbation of the ``smooth'' case, that is, the case where these local sharp features are not taken into account. In order to make a comparison with these previous results, we thus need to express the eigenfrequency deviations induced by low-amplitude glitches that is predicted by the present framework, while additionally using the small perturbations hypothesis. This is actually equivalent to assuming that (a) the barriers are very thin and localized at a frequency-independent radius, namely, $r_{i+1}^-(\sigma)=r_i^+(\sigma)=r_{{\rm g},i}$; and (b) the barriers are weakly reflective, that is, $|R_{i,i+1}|\ll 1$. The goal is then to find the frequency perturbation $\sigma_1$ induced by the barrier, that is,
\algn{
\sigma_1=\sigma-\sigma_0 \; ,
}
where $\sigma$ is the actual mode angular eigenfrequency and $\sigma_0$ indicates its value in the smooth case without any barrier .

\subsubsection{One single glitch}

We first consider the most familiar case of one single glitch located between two resonant cavities $C_1$ and $C_{2}$. Within the perturbation framework, the phase terms in \eq{Ups_i} can be rewritten in any cavity $C_i$ as
\algn{
\Upsilon_i(\sigma)=\Upsilon_i^{(0)}(\sigma)+\Upsilon_i^{(1)}(\sigma)\; ,
\label{decompose Y}
}
where the superscript $(1)$ indicates in the following a perturbation induced by the barriers whereas the superscript $(0)$ indicates a value (or any structural function) in the smooth case when the barrier is not taken into account.
Neglecting the perturbation of the wave number integral function because of the hypothesis (a) and making use of the fact that the core and surface phase lags are not perturbed by the barriers, we can deduce for the two-cavity case that
\algn{
\Upsilon_1^{(0)}(\sigma)&=\Theta_1^{(0)}(\sigma)+\frac{\delta_{\rm c}^{(0)}(\sigma)}{2} \label{Ups^(0)_1}\\
\Upsilon_2^{(0)}(\sigma)&=\Theta_2^{(0)}(\sigma)+\frac{\delta_{\rm s}^{(0)}(\sigma)}{2} \label{Ups^(0)_N} \; ,
}
and
\algn{
\Upsilon_1^{(1)}(\sigma)&=\frac{\delta_{1,2}^{(1)}(\sigma)}{2}-\frac{\pi}{2}\label{Ups_1^(1)}\\
\Upsilon_2^{(1)}(\sigma)&=\frac{\delta_{2,1}^{(1)}(\sigma)}{2}\label{Ups_2^(1)}\; ,
}
which is consistent with the definitions in \eqsduo{Ups_i}{Ups_1}.
First, in the ``smooth'' configuration without any barrier, the frequency condition is similar to \eq{1 cavity} in the case of one single cavity located between the center and surface of the star, that is:
\algn{
\overline{\Upsilon}_2^{(0)}(\sigma_0)= n \pi\; ,
\label{smooth cond 2}
}
where $n$ is the radial order and where the overline notation for any integers $i$ and $k$ is defined as
\algn{
\overline{\Upsilon}_i^{(k)}=\sum_{j=1}^{i} \Upsilon_j^{(k)} \; .
}
Second, adding the intermediate glitch, we choose to write the general resonance condition in \eqsduo{tan Phi}{cond tan Phi}, or equivalently in \eq{resonance math},~as a series of sine terms as in \eq{big resonance}, that is, for $N=2:$
\algn{
\sin\left[ \Upsilon_1(\sigma)+\Upsilon_2(\sigma)\right]=-\left|R_{1,2}(\sigma)\right|\sin\left[ \Upsilon_2(\sigma)-\Upsilon_1(\sigma)\right] \label{cond glitch}\; .
}
We emphasize that this resonance condition is also valid to describe mixed modes, except that in the present case, the modes have the same nature (i.e., gravity or acoustic) on both sides of the intermediate barrier (and thus the same sign in front of the wave number integral in both cavities). Within the small perturbation limit, we assume that the frequency perturbation $\sigma_1$ is small enough that the perturbations of the reflection coefficient and of the total ``optical'' path of the waves during one travel through the ensemble of cavities is also small compared to the smooth case, that is,
\algn{
\left( \deriv{\ln| R_{1,2}|}{\sigma}\right)_{\sigma_0}\sigma_1\ll 1,\;\overline{\Upsilon}_2^{(1)}(\sigma) \ll \pi,\; \left(\deriv{\Upsilon_i^{(k)}}{\sigma}\right)_{\sigma_0}\sigma_1  \ll \pi \; .
\label{small phase 2}
}
Using \eqsduo{decompose Y}{smooth cond 2}, a first-order Taylor expansion of \eq{cond glitch} in $|R_{1,2}|$ by hypothesis (b) and in the quantities in \eq{small phase 2} leads to
\algn{
\frac{\pi\sigma_1}{\Delta \sigma}  +\overline{\Upsilon}_2^{(1)}(\sigma_0)\approx
\left|R_{1,2}(\sigma_0)\right| \sin \left(2\left[\Upsilon_1^{(0)}(\sigma_0)+\Upsilon_1^{(1)}(\sigma_0)\right] \right) \; ,
\label{multi g b}
}
where we have defined the ``pseudo'' pulsation large separation in the general case of $N$ cavities as
\algn{
\Delta \sigma=\pi\left(\deriv{\overline{\Upsilon}_N^{(0)}}{\sigma}\right)_{\sigma_0}^{-1} \; .
}
Using \eqs{base adjoint relations}{Ups_1^(1)}{Ups_2^(1)}, \eq{multi g b} can be rewritten as
\algn{
\frac{\sigma_1}{\Delta \sigma}\approx -\frac{\left|R_{1,2}(\sigma_0)\right|}{\pi} \sin \left(2\Upsilon_1^{(0)}(\sigma_0)+\delta_{1,2}(\sigma_0) \right) -\frac{\gamma_{1,2}^{(1)}(\sigma_0)}{\pi} \; .
\label{multi g b 2}
}
Moreover, using \eqsduo{Theta_1}{Ups^(0)_1}, we obtain:~
\algn{
\Upsilon_1^{(0)}(\sigma_0)&=\pm \int_{r_1^-\left(\sigma_0\right)}^{r_{\rm g,1}} \mathcal{K}_r\left(\sigma_0\right)\dd r+\frac{\delta_{\rm c}^{(0)}\left(\sigma_0\right)}{2} \label{acoustic radius} \; .
}
The frequency deviation induced by the low-amplitude glitch from the smooth case therefore takes the form of a small offset term plus a sinusoidal term. The magnitude of the sinusoidal perturbation is proportional to the reflection coefficient associated with the glitch. Moreover, its argument depends on the wave number integral from the core to the radius of the glitch, and is related to the so-called acoustic radius. The phase offset results from the phase lag introduced at reflection on the glitch. We note that it is also possible to formulate \eq{multi g b 2} as a function of \smash{$\Upsilon_2^{(0)}(\sigma_0)$ instead of $\Upsilon_1^{(0)}(\sigma_0)$} using \eq{smooth cond 2}, that is as a function of the wave number integral from the glitch radius to the surface, which is related to the so-called acoustic depth. 

The description of the first-order effect of low-amplitude glitches on the eigenfrequencies that is provided by \eq{multi g b 2} finally appears to be consistent with the computation of \cite{Pincon2019b}. It is also compatible with previous works based on the variational principle (see \sectionname{}~\ref{introduction} for references);  in addition, it has the advantage of clarifying the general physical meaning of the different terms inside the glitch-induced deviation (e.g., amplitude, offsets). While the approach based on the variational principle is limited to the first-order small-amplitude glitches only, we stress that the formulation developed in this paper is more general and can also be used to treat big glitches (e.g., mode-trapping phenomena).

\subsubsection{Multiple glitches}

 In a second step, it is straightforward to extend the last first-order expansion to the case of $N-1$ low-amplitude glitches. Using the hypothesis (a) and the non-perturbation of the core and surface phase lags, the phase terms in \eq{decompose Y} can be expressed for $1\le i \le N$ within the perturbation framework as:
\algn{
\Upsilon_i^{(0)}(\sigma)&=\Theta_i^{(0)}(\sigma)+\delta_{\rm K}(i-1)\frac{\delta_{\rm c}^{(0)}(\sigma)}{2}+\delta_{\rm K}(i-N)\frac{\delta_{\rm s}^{(0)}(\sigma)}{2} \label{Ups^(0)},\\
\Upsilon_i^{(1)}(\sigma)&=\frac{\delta_{i,i+1}^{(1)}(\sigma)}{2}+\frac{\delta_{i,i-1}^{(1)}(\sigma)}{2} -\frac{\pi}{2} \label{Ups^(1)}\; ,
}
where $\delta_{\rm K}(i)$ is the Kronecker delta function and we set
\algn{
\delta_{1,0}^{(1)}=0~~~~\mbox{and}~~~~\delta_{N,N+1}^{(1)}=\pi \label{delta^(1)} 
}
in order to be consistent with the definitions in \eqsduo{Ups_i}{Ups_1}.
First, in the ``smooth'' configuration without any barrier, the frequency condition is expected to be similar to \eq{smooth cond 2} but for $N$ cavities, that is,
\algn{
\overline{\Upsilon}_N^{(0)}(\sigma_0)= n \pi\; .
\label{smooth cond}
}
Second, we add the $N-1$ low-amplitude glitches. Within the perturbation limit, we assume as previously that the frequency perturbation is small enough that:
\algn{
\left( \deriv{\ln |R_{i,i+1}|}{\sigma}\right)_{\sigma_0}\sigma_1\ll 1,\;\overline{\Upsilon}_N^{(1)}(\sigma) \ll \pi,\; \left(\deriv{\Upsilon_i^{(k)}}{\sigma}\right)_{\sigma_0}\sigma_1  \ll \pi \; .
\label{small phase}
}
Using the resonance condition in \eq{big resonance} for $N$ cavities, a first-order Taylor expansion as a function of $|R_{i,i+1}|$ by hypothesis (a) and of the small quantities in \eq{small phase} leads to:
\algn{
\frac{\sigma_1}{\Delta \sigma}\approx \sum_{i=1}^{N-1} \frac{\left|R_{i,i+1}(\sigma_0)\right|}{\pi} \sin \left(2\overline{\Upsilon}_i^{(0)}(\sigma_0)+2\overline{\Upsilon}_i^{(1)}(\sigma_0) \right) -\frac{\overline{\Upsilon}_N^{(1)}(\sigma_0)}{\pi} \; .
\label{multi g 2}
}
Moreover, using \eqsduo{Theta_1}{Ups^(0)}, and hypothesis (a) on the one hand, and \eqs{base adjoint relations}{Ups^(1)}{delta^(1)}~on the other hand, we can deduce for $i\le N-1$ that:
\algn{
\overline{\Upsilon}_i^{(0)}(\sigma_0)&=\pm \int_{r_1^-(\sigma_0)}^{r_{{\rm g},i}(\sigma_0)} \mathcal{K}_r(\sigma_0)\dd r+\frac{\delta_{\rm c}^{(0)}(\sigma_0)}{2} \\
\overline{\Upsilon}_i^{(1)}(\sigma_0)&=\sum_{j=1}^{i-1}\gamma_{j,j+1}^{(1)}(\sigma_0)+\frac{\delta_{i,i+1}^{(1)}(\sigma_0)}{2}-\frac{\pi}{2}\; .
}
The total frequency deviation from the smooth case therefore reduces to the superposition of the perturbations resulting from each glitch, which are related to the acoustic radius and the phase lags at transmission and reflection.

%
\section{Concluding remarks and discussion}
\label{conclusion}

 In this work, we derive a general analytical expression for the asymptotic resonance condition of global oscillation modes in spherical stars. While this can seem complex at first glance, we show that this is, in fact, merely analogous to regard a star as a one-dimensional giant Fabry-Pérot interferometer, composed of a multitude of resonant cavities and impermeable surface and core boundaries. In the adopted view, a star is decomposed into an ensemble of resonant cavities where waves can propagate within the short-wavelength JWKB approximation. The cavities are separated by intermediate barriers corresponding to evanescent regions or glitches. Each barrier is associated with a reflection and transmission coefficient. The core and surface boundary conditions are represented by totally reflective barriers. Within this framework, we obtained the resonance condition while considering two different physical pictures. In the infinite-time reflection picture, we follow the back-and-forth travel of a wave energy ray through the ensemble cavities and assume constructive interferences; the resonance condition is provided by \eqsduo{tan Phi}{cond tan Phi}. In the second picture, the eigenmodes are considered as the solution of a linear boundary value problem and the resonance condition is provided by \eq{resonance math}. Both pictures are equivalent and predict the same resonance condition, which turns out to depend on a number of parameters: the wave number integral over each cavity; the coupling factor associated with each barrier between adjacent cavities; the phase lags at reflection and transmission through the stars.
In addition, the amplitude ratio between adjacent cavities is also expressed analytically as a function of these parameters in \sectionname{}~\ref{mode energy and amplitude} and can inform us on the distribution of the mode energy through the star. The present formulation enables us to retrieve in a convenient way simple cases already widely studied in the past, as one single cavity acoustic or gravity mode or two- and three-cavity mixed modes as well as the case of multiple low-amplitude glitches.
 
This general formulation provides a physically grounded interpretation of the mode resonance condition in stars. It also provides a useful tool for analyzing, interpreting, or predicting the oscillation power spectra in a practical way and across a broad variety of configurations and evolutionary stages. This new diagnosis tool is for instance expected to bring a valuable help for the development of automated seismic analysis pipelines, which are needed in the current era of space missions and especially for the future PLATO mission \citep[e.g.,][]{Rauer2014}. The next step in the investigation will consist in expressing the wave reflection and transmission coefficients around the barriers for specific cases using either toy models or more realistic stellar structure models. By means of either analytical or numerical methods to solve the wave equation, such a study can allow for a link to be forged between the mode parameters, the mode frequencies, and the stellar interior properties. This will be further developed in future works and practical applications on observed spectra will be undertaken.

Finally, we also note that the present general formulation can be applied to study the transmission of progressive waves through layered media in stellar interiors or planetary atmospheres since it offers the possibility to easily retain and remove  the boundary conditions according to the configuration. This can be useful in addressing the problem of the transport of angular momentum, heat, or chemical elements by waves in these objects and the present results provide a general formalism for tackling such issues \citep[e.g.,][]{Andre2019,Cai2021}

\begin{acknowledgements}
During this work, C. P. was funded by a postdoctoral fellowship of Chargé de Recherche from F.R.S.-FNRS (Belgium). C. P. also acknowledges financial support from Sorbonne Université (EMERGENCE project). M. T. is partially supported by JSPS KAKENHI Grant Number 18K03695.
\end{acknowledgements}

\bibliographystyle{aa} 
\bibliography{bib}


\begin{appendix}

\section{Gravito-acoustic wave equations in the Cowling approximation}
\label{Shiba}

In this section, we briefly recall the form of the second-order equations for gravito-acoustic waves as formulated by \cite{Shibahashi1979} under the Cowling approximation. 

A first version is provided by \eqss{wave equation}{K_r JWKB}~where the $M$ function is equal to
\algn{
M(r)&= f\left[P(r)\right] \; ,
}
where
\algn{
f(y)&=H_p^2\left| y\right|^{1/2} \derivs{\left| y\right|^{-1/2}}{r}=-\frac{H_p^2}{2}\derivs{\ln \left| y \right|}{r}+\frac{H_p^2}{4}\left( \deriv{\ln \left| y \right|}{r}\right)^2  \\
P(r)&=\frac{r^2 h(r)}{c^2} \left(\frac{S_\ell^2}{\sigma^2}-1 \right) \\
h(r)&=\exp\left[ \int_0^r\left(\frac{N^2}{g}-\frac{g}{c^2} \right) \dd r\right] \; .
}
We also have the  following relations linking the radial displacement with the perturbation of pressure denoted by $\tilde{p}^\prime(r)$ in \eq{p pert} \citep[e.g.,][]{Unno1989},
\algn{
\widetilde{\Psi}&=\rho^{-1/2} r \left | N^2-\sigma^2\right |^{-1/2} \tilde{p}^\prime e^{-\di \sigma t} \label{var w}\\
\widetilde{\Psi}&={\rm sgn}(P) \frac{1}{|\mathcal{K}_r|} \left( \deriv{\Psi}{r}+\frac{1}{2} \deriv{\ln |P|}{r}~ \Psi\right) \label{p by xi}\; ,
}
where sgn() is the sign function. We note that the $M$ function is on the order of unity at most over regions that are far away from the turning points where $P=0$ (i.e., where $\sigma^2=S_\ell^2$) or from the sharp gradients in the structure.

In the vicinity of a turning point where $P=0$, the first version of the wave equation in \eq{wave equation} is singular.
Around such a turning point, it is therefore more convenient to use the second version of the wave equation, which is provided by
\algn{
&\derivs{\widetilde{\Psi}}{r} + \left[\mathcal{K}_r^2 -\frac{L(r)}{H_p^2}\right]\widetilde{\Psi} = 0 \; ,
\label{wave equation 2}
}
where
\algn{
&L(r)=f\left[Q(r)\right] \\
&Q(r)=\frac{1}{r^2 h(r)}(\sigma^2-N^2) \; .
}
The link between the perturbation of pressure and the radial displacement can in this case be obtained through the equation
\algn{
\Psi&={\rm sgn}(Q) \frac{1}{|\mathcal{K}_r|} \left( \deriv{\widetilde{\Psi}}{r}+\frac{1}{2} \deriv{\ln |Q|}{r}~ \widetilde{\Psi}\right)\; .
\label{Psi by Psi_tilde}
}
The $L$ function is on the order of unity at most over regions that are far away from the turning points where $Q=0$ (i.e., where $\sigma^2=N^2$) or from the sharp gradients in the structure. Moreover, at turning points such as $Q=0$, this second version of the wave equation is conversely singular, and it is therefore more convenient to consider the first version of the wave equation in \eq{wave equation}.
 
\section{Relation between the base and adjoint wave transmission-reflection coefficients}
\label{Takata}

As shown by \cite{Takata2016b}, the reflection and transmission coefficients in a base wave transmission-reflection problem are related to those of the adjoint wave transmission-reflection problem (see \sectionname{}~\ref{barrier} for a description of the two configurations). In this section, we recall the result obtained in Sect.~2 of \cite{Takata2016b} and generalize it to the case of a complex transmission coefficient (i.e., accounting for a transmission phase lag).

In the base wave reflection-transmission problem, the global wave function $\Psi(r,t)$ is provided in the cavity $C_n$, within the convention considered in \sectionname{}~\ref{barrier} and up to a proportionality constant, by
\algn{
\Psi_n(r,t;r_n^+)=\psi^{(\rightarrow)}(r,t;r_n^+)~+~R_{n,n+1}~\psi^{(\leftarrow)}(r,t;r_n^+) \; ,
\label{sol 1}
}
while in the overlying cavity $C_{n+1}$, it is written
\algn{
\Psi_{n+1}(r,t;r_{n+1}^-)=T_{n,n+1}~\psi^{(\rightarrow)}(r,t;r_{n+1}^-) \; .
\label{sol 2}
}
The goal is then to build a solution of the adjoint problem from the two representations of the global wave function $\Psi(r,t)$ in \eqsduo{sol 1}{sol 2}. To do so, we can use the fact that the oscillation equations are invariant by the operations of (a) time reversal and (b) complex conjugation. As a justification, (a) the wave equation introduced in \eqss{wave equation}{K_r JWKB} and \appendixname{}~\ref{Shiba} is independent of the sign of $\sigma$, and (b) its coefficients are real. Therefore, if $\Psi(r,t)$ is a solution, then $\Psi^\prime(r,t)=\Psi^\star(r,-t)$ is also a solution. 
According to \eq{Phi transf}, applying the operations of time reversal and complex conjugation on \eqsduo{sol 1}{sol 2}~provides the form of the new solution $\Psi^\prime(r,t)$, which can be respectively expressed in the cavities $C_n$ and $C_{n+1}$ as
\algn{
\Psi_n^\prime(r,t;r_n^+)&=R_{n,n+1}^\star\psi^{(\rightarrow)}(r,t;r_n^+)~+~\psi^{(\leftarrow)}(r,t;r_n^+) \label{sol 1 p}\\
\Psi_{n+1}^\prime(r,t;r_{n+1}^-)&=T_{n,n+1}^\star~\psi^{(\leftarrow)}(r,t;r_{n+1}^-) \label{sol 2 p}\; .
}
Another solution can then be built from a linear combination of the solution $\Psi$ represented by \eqss{sol 1}{sol 2} and the solution $\Psi^\prime$ represented by \eqss{sol 1 p}{sol 2 p}. This must be done in such a way that the upward component in the cavity $C_n$ vanishes in order to retrieve the adjoint problem. Such a solution $\Psi^{\prime\prime}$ can be obtained by the following linear combination
\algn{
\Psi^{\prime\prime}(r,t)=e^{2\di \gamma_{n,n+1}}\frac{\left[-R_{n,n+1}^\star\Psi(r,t)+\Psi^\prime(r,t) \right]}{T_{n,n+1}} \; ,
}
which must be simultaneously  applied in the cavity $C_n$, that is,
\algn{
e^{2\di \gamma_{n,n+1}}\frac{\left[-R_{n,n+1}^\star\Psi_n(r,t;r_n^+)+\Psi_n^\prime(r,t;r_n^+) \right]}{T_{n,n+1}}=T_{n,n+1} \psi^{(\leftarrow)}(r,t;r_n^+) \label{sol 1 c} \; ,
}
and in the cavity $C_{n+1}$, that is,
\algn{
e^{2\di \gamma_{n,n+1}}&\frac{\left[-R_{n,n+1}^\star\Psi_{n+1}(r,t;r_{n+1}^-)+\Psi_{n+1}^\prime(r,t;r_{n+1}^-) \right]}{T_{n,n+1}} \nonumber\\
&=-R_{n,n+1}^\star e^{2\di \gamma_{n,n+1}}~\psi^{(\rightarrow)}(r,t;r_{n+1}^-)+\psi^{(\leftarrow)}(r,t;r_{n+1}^-) \label{sol 2 c} \; .
}
By comparing \eq{sol 1 c} with \eq{Phi sol n} and \eq{sol 2 c} with \eq{Phi sol n+1}, and using the definitions in \eqsduo{R_n+1_n}{T_n+1_n}, we can conclude by identification:
\algn{
T_{n+1,n}=T_{n,n+1}~~\mbox{and}~~R_{n+1,n}=|R_{n,n+1}|~e^{\di\left(\pi-\delta_{n,n+1}+2\gamma_{n,n+1}\right)} \; .
}
%

\section{Alternative formulation for the resonance condition}
\label{sine}

Depending on the problem, it can be better to express the resonance condition as a series of sine terms that are functions of the $\Upsilon_i$ phases in \eq{Ups_i}. To do so, we first rewrite, for sake of the convenience, the resonance condition provided in \eqsduo{tan Phi}{cond tan Phi}~in such a way to isolate the recurrence variable on the level $i$ rather than on the level $i-1$.
This can be done in a simple way by defining $\tau_i=\arctan\left[q_{i-1,i} \cot \Phi_{i-1} \right]$, so that
\eqsduo{tan Phi}{cond tan Phi}~are equivalent, for $i\ge 2,$ to the conditions
 \algn{
 \tan \tau_i = - q_{i-1,i} \tan\left( \Upsilon_{i-1}-\tau_{i-1}\right) \; ,~ \tau_1&=0 \; ~\mbox{and}\; ~\tau_N=\Upsilon_N \; .
 \label{tan tau}
 }
Second, we can proceed step by step. Using \eq{coupling} and usual trigonometric formula, \eq{tan tau} for $i=N$ provides
\algn{
\sin \left( \Upsilon_N +\Upsilon_{N-1}-\tau_{N-1}\right)+\left| R_{N-1,N}\right|\sin \left(\Upsilon_N-\Upsilon_{N-1}+\tau_{N-1} \right)=0 \; .
\label{form N}
}
Then, using \eq{tan tau} to express $\tau_{N-1}$ as a function of $\tau_{N-2}$ and using the identities $\cos (\arctan x)=1/(1+x^2)^{1/2}$ and $\sin (\arctan x)=x/(1+x^2)^{1/2}$, we can express \eq{form N} as
\algn{
A_N +\left| R_{N-1,N} \right|B_N=0 \; ,
}
with
\algn{
&A_N \propto \sin \left( \Upsilon_N +\Upsilon_{N-1}+\Upsilon_{N-2}-\tau_{N-2}\right)\nonumber\\
&~~~~+\left|R_{N-2,N-1}\right| \sin \left( \Upsilon_N +\Upsilon_{N-1}-\Upsilon_{N-2}+\tau_{N-2}\right) \label{A_N} \\
&B_N \propto\sin \left( \Upsilon_N -\Upsilon_{N-1}-\Upsilon_{N-2}+\tau_{N-2}\right)\nonumber\\
&~~~~+\left|R_{N-2,N-1}\right| \sin \left( \Upsilon_N -\Upsilon_{N-1}+\Upsilon_{N-2}-\tau_{N-2}\right) \label{B_N}\; ,
}
where the proportionality constant is the same for $A_N$ and $B_N$.
We note that the expressions of $A_N$ and $B_N$ are very similar to the left-hand side of \eq{form N}. Successively performing  the same operation as before, that is, using \eq{tan tau} to express $\tau_{N-2}$ as a function of $\tau_{N-3}$, we can express \eqsduo{A_N}{B_N}~as
\algn{
A_N &\propto A_{N-1}^{(1)}+\left|R_{N-2,N-1}\right|B_{N-1}^{(1)}\\
B_N &\propto A_{N-1}^{(2)}+\left|R_{N-2,N-1}\right|B_{N-1}^{(2)} \; ,
}
where
\algn{
&A_{N-1}^{(1)} = \sin \left( \Upsilon_N +\Upsilon_{N-1}+\Upsilon_{N-2}+\Upsilon_{N-3}-\tau_{N-3}\right)\nonumber\\
&~~~~+\left|R_{N-3,N-2}\right| \sin \left( \Upsilon_N +\Upsilon_{N-1}+\Upsilon_{N-2}-\Upsilon_{N-3}+\tau_{N-3}\right) \\
&B_{N-1}^{(1)} = \sin \left( \Upsilon_N +\Upsilon_{N-1}-\Upsilon_{N-2}-\Upsilon_{N-3}+\tau_{N-3}\right)\nonumber\\
&~~~~+\left|R_{N-3,N-2}\right| \sin \left(\Upsilon_N +\Upsilon_{N-1}-\Upsilon_{N-2}+\Upsilon_{N-3}-\tau_{N-3}\right) \\
&A_{N-1}^{(2)} = \sin \left( \Upsilon_N -\Upsilon_{N-1}-\Upsilon_{N-2}-\Upsilon_{N-3}+\tau_{N-3}\right)\nonumber\\
&~~~~+\left|R_{N-3,N-2}\right| \sin \left( \Upsilon_N -\Upsilon_{N-1}-\Upsilon_{N-2}+\Upsilon_{N-3}-\tau_{N-3}\right)\\
&B_{N-1}^{(2)} = \sin \left( \Upsilon_N -\Upsilon_{N-1}+\Upsilon_{N-2}+\Upsilon_{N-3}-\tau_{N-3}\right)\nonumber\\
&~~~~+\left|R_{N-3,N-2}\right| \sin \left( \Upsilon_N -\Upsilon_{N-1}+\Upsilon_{N-2}-\Upsilon_{N-3}+\tau_{N-3}\right)
}
The final expression can therefore be obtained by performing the same operation several times until reaching the variable $\tau_1=0$.
At the end of the day, the compact general form of the resonance condition reads
\algn{
&\sin \left(\sum_{j=1}^{N}\Upsilon_j \right)
+\sum_{i=1}^{N-1} \left\{ \sum_{(p_k)\in E_i}\left( \prod_{k=1}^{i} \left|R_{p_k,p_k+1}\right| \right) \sin \left(\sum_{j=1}^{N} c_j \Upsilon_j \right)\right\}=0 \; ,
\label{big resonance}
}
where $E_i$ represents the set of the arrangements of $i$ distinct elements in $I_{N-1}=\llbracket 1,N-1 \rrbracket$ listed in descending order, that is, $E_i=\{(p_k)_{1\le k \le i}\}$ such as $p_k\in I_{N-1}$ and $p_{k+1}<p_k$. The coefficients $\{c_j\}_{1\le j \le N}$ in the sine function are equal to either $c_j=-1$ if the $k$ index of the lowest $p_k$ value such as $p_k \ge j$ is odd or $c_j=+1$ otherwise (i.e., $c_N$ is always equal to $+1$). We note that each set $(E_i)_{i\in I_{N-1}}$ is composed of $C(N-1,i)$ arrangements, where $C(N-1,i)$ are the usual binomial coefficients. The number of terms in \eq{big resonance} is thus equal the sum of the number of arrangements in each set $(E_i)_{i\in I_{N-1}}$ plus one for the first term, which results in $2^{N-1}$ terms.

\section{Local mean mode energy in cavities}
\label{energy}

We detail in this section the calculation of \eq{local energy} within the asymptotic limit. First, using  \eqsduo{v_r xi}{v_r}, we can get the JWKB form of the radial displacement inside a given cavity, that is,
\algn{
\tilde{\xi}_r=\left( \frac{1}{\rho r^2 c\sigma}\right)^{1/2} \left| \frac{S_\ell^2-\sigma^2}{N^2-\sigma^2}\right| ^{1/4}  \left( a_{\rm p}~e^{i\varphi} + a_{\rm r}~ e^{-i\varphi}\right) 
\; . \label{xi_r app}
}
Using \eqsduo{p^prime}{xi_h}, we can also get the JWKB form of the horizontal displacement function, that is,
\algn{
\tilde{\xi}_h={\rm sgn}\left(S_\ell^2-\sigma^2\right) \di \left(\frac{c}{ \rho r^4 \sigma^3}\right)^{1/2} \left| \frac{N^2-\sigma^2}{S_\ell^2-\sigma^2}\right| ^{1/4}  \left( a_{\rm p}~e^{i\varphi} - a_{\rm r}~ e^{-i\varphi}\right) 
\; . \label{xi_h app}
}
Second, we recall that the orthonormal spherical harmonics are given by
\algn{
Y_\ell^m(\theta,s)&=N_{\ell m}~P_\ell^m(\cos \theta)~ e^{\di ms} \; , \label{SH}
}
where $P_\ell^m$ are the associated Legendre polynomials. The normalization constant is given by
\algn{
N_{\ell m}&=\sqrt{\frac{(2\ell +1)}{4\pi}\frac{(\ell-m)!}{(\ell+m)!}} \; ,
}
and is such that
\algn{
&\int_0^\pi \left[P_\ell^{m}(\cos \theta)\right]^2 \sin \theta \dd \theta = \frac{1}{2\pi N_{\ell m}^2}\; .
\label{norm}
}
We also recall that the spherical harmonics are solutions of the eigenvalue equation $r^2\vec{\nabla}^2 Y_\ell^m=-\ell(\ell+1) Y_\ell^m$. Multiplying this eigenvalue equation by the complex conjugate of $Y_\ell^m$, integrating over the sphere, using \eq{norm} for the right-hand side of the equations and integration by parts for the left-hand side, we can retrieve the well-known relation
\algn{
&\int_0^\pi   \left[\left(\deriv{}{\theta}\left[P_\ell^m(\cos \theta)\right]\right)^2+\frac{m^2}{\sin^2\theta} \left[P_\ell^m(\cos \theta)\right]^2\right] \sin \theta \dd \theta  =\frac{\ell(\ell+1)}{2 \pi N_{\ell m}^2} \; .
\label{norm 2}
}
Finally, taking the square modulus of the real part of \eq{xi pert} while considering the JWKB form of the displacement and the spherical harmonics expression in \eqss{xi_r app}{SH}, integrating first on $t$ and $s$, and then integrating over $\theta$ using \eqsduo{norm}{norm 2}, \eq{local energy} can be rewritten to a good approximation as
\algn{
\mathcal{E}_k&\approx\frac{|a_{\rm p}|^2+|a_{\rm r}|^2}{2}\int_{r_k^-}^{r_k^+} \mathcal{K}_r \left(\left|\frac{N^2}{\sigma^2}-1 \right|^{-1}+\left|1-\frac{\sigma^2}{S_\ell^2} \right|^{-1} \right)\dd r \; , \label{E_k complete}
}
where $\mathcal{K}_r$ is defined in \eq{K_r JWKB} and where we have also used the fact that the wave number integral is much higher than unity within the asymptotic limit to neglect the residual oscillating radial functions in the integrand. Either in the case of pressure-dominated modes, namely, \smash{$\sigma^2 \gg (S_\ell^2~\mbox{and}~N^2)$}, or gravity-dominated modes, namely, \smash{$\sigma^2 \ll (S_\ell^2~\mbox{and}~N^2)$}, the term in brackets in the integrand of \eq{E_k complete} is close to unity. Moreover, as noted in \sectionname{}~\ref{surface boundary}, $|a_{\rm p}|=|a_{\rm r}|=|a_k|$ inside the cavity $C_k$ in the case of eigenmodes with totally reflective boundary conditions. This justifies the validity of \eq{energy approx}.

\section{Supplementary expressions of the amplitude ratios}
\label{local-to-core}

In \sectionname{}~\ref{surface boundary}, we use the surface boundary condition to express the mode amplitude in the cavity $C_k$ as a function of that in the upper cavity $C_N$. Imposing instead the core boundary condition, we aim in this section to express $a_k$ as a function of $a_1$ in the inner cavity $C_1$. 
To do so, we first invert the relation in \eq{general transf} to obtain for $1\le i \le N-1$
\algn{
\vec{a}_{i+1,+}&\equiv\mathsf{E}_{i+1,i} ~\vec{a}_{i,+}
\label{connection 2}
}
where
\begingroup\addtolength{\jot}{0.2cm}
\algn{
\mathsf{E}_{i+1,i}&\equiv\mathsf{E}_{i,i+1}^{-1} = \mathsf{C}\left(\Theta_{i+1} \right)\mathsf{C}\left(\frac{\delta_{i+1,i}}{2}-\frac{\pi}{2} \right)~ \mathsf{A}_{i+1,i} ~\mathsf{C}\left(\frac{\delta_{i,i+1}}{2}\right) \\
\mathsf{A}_{i+1,i}&\equiv\mathsf{A}_{i,i+1}^{-1}= \frac{1}{\left|T_{i,i+1}\right|}\left(
\begin{array}{cc}
1&-\left|R_{i,i+1}\right|\\
-\left|R_{i,i+1}\right|&1
\end{array} \right) \; ,
}
\endgroup
with $\mathsf{C}(\varepsilon)$ and $\mathsf{A}_{i,i+1}$ defined in \eqsduo{C matrix}{A matrix}. Using \eq{C transf}, it is then straightforward to show that the application of $\mathsf{E}_{i+1,i}$ on a vector in the form of \eq{w vector} leads to
\algn{
\mathsf{E}_{i+1,i}~\vec{\varw}(\varepsilon)=\mathcal{A}_{i+1,i}\left(\varepsilon+\frac{\delta_{i,i+1}}{2}\right)~\vec{\varw}\left[ \varOmega_{i+1}^{-1}(\varepsilon)\right] \; ,
\label{A transf 2}
}
where
\algn{
\mathcal{A}_{i+1,i} (\varepsilon)
&= \sqrt{\frac{1+(q_{i,i+1}^2-1)\cos^2 \left( \varepsilon \right)}{q_{i,i+1}}}\; ,
\label{A amplitude 2}
}
and in which we have used the convention for the branch of the $\arctan$ function in \eq{branch}. Within this convention, $\varOmega_{i+1}^{-1}$ is the inverse function of $\varOmega_{i+1}$ in \eq{Omega} and is defined for $1\le i\le N-1$ as
\algn{
\varOmega_{i+1}^{-1}(\varepsilon)=\arctan\left[ \frac{1}{q_{i,i+1}}\tan\left(\varepsilon+\frac{\delta_{i,i+1}}{2} \right)\right]+\Upsilon_{i+1}-\frac{\delta_{i+1,i+2}}{2}\; .
}
Second, using the inverse of \eq{matching}, the core boundary condition implies that
\algn{
\vec{a}_{1,+} = \mathsf{C}(\Theta_1)~a_1~ \vec{\varw}\left(\frac{\delta_{\rm c}}{2} \right)\; , \label{a_1 boundary 2}
 }
Finally, applying $k-1$ times the transformation in \eq{connection 2} on the upper core amplitude vector in \eq{a_1 boundary 2}, we can obtain the amplitude vector in the cavity $C_k$ for $k>1$, that is,
\algn{
\vec{a}_{k,+}&=\left(\prod_{i=k-1}^{1} \mathsf{E}_{i+1,i} \right) \mathsf{C}(\Theta_1)~a_1~\vec{\varw}\left(\frac{\delta_{\rm c}}{2} \right) \label{a_k prod up} \; .
}
Using the properties in \eqsduo{C transf}{A transf 2}, \eq{a_k prod up} can be written for $k>1$ as
\algn{
\vec{a}_{k,+}=  a_k ~ \vec{\varw} \left[  \varOmega^{-1}_{k} \circ \cdot \cdot \cdot \circ \varOmega^{-1}_1 \left(0\right) \right] \; , \label{a_k prod 2 up}
}
where $(\circ)$ is the composition operator, $\varOmega_1^{-1}$ is the inverse of the function in \eq{Omega_1}, and $a_k$ is given by
\algn{
\frac{a_k}{a_1}=\prod_{i=k-1}^{1}  \mathcal{A}_{i+1,i} \left[\varOmega^{-1}_{i} \circ \cdot \cdot \cdot \circ \varOmega^{-1}_1 \left(0\right)+\frac{\delta_{i,i+1}}{2}\right] \; .
\label{a_k/a_1}
}
According to \eqsduo{A amplitude 2}{a_k/a_1}, we can therefore deduce that the squared amplitude ratio between the adjacent cavities $C_k$ and $C_{k+1}$ is equal to
\algn{
 \left|\frac{a_k}{a_{k+1}}\right|^2= \dfrac{q_{k,k+1}}{q_{k,k+1}^2+(1-q_{k,k+1}^2) \cos^2\left( \Xi^\prime_k \right)} \; , \label{amp ratio 2}
}
where we have defined
 \algn{
 \Xi_k^\prime&=\varOmega^{-1}_{k} \circ \cdot \cdot \cdot \circ \varOmega^{-1}_{1} \left(0\right)+\frac{\delta_{k,k+1}}{2}+\frac{\pi}{2}\; ,
 }
which measures the level of constructive interference in the cavity $C_k$. The same discussion as that based on \eq{amp ratio} in \sectionname{}~\ref{ratio} holds true here when only the core boundary condition is met, except that $\Xi_{k+1}$ and $C_{k+1}$ have to be replaced by $\Xi_k^\prime$ and $C_k$.
We insist on the fact that \eqsduo{amp ratio}{amp ratio 2}~are strictly equivalent only within the resonance condition, that is, when the core and surface boundary conditions are simultaneously met. Under this condition, we note that we can also obtain another equivalent relation by multiplying both equations, which is reduced to:
\algn{
 \left|\frac{a_k}{a_{k+1}}\right|^4= \dfrac{q_{k,k+1}^2+(1-q_{k,k+1}^2) \cos^2\left( \Xi_{k+1} \right)}{q_{k,k+1}^2+(1-q_{k,k+1}^2) \cos^2\left( \Xi^\prime_{k} \right)} \; .
 \label{amp 4}
}
We notice that Eqs~(\ref{amp ratio}), (\ref{amp ratio 2}) and (\ref{amp 4}) generalize the expressions of the amplitude ratios derived in Sect. 3.2 of \cite{Takata2016b} in the case of mixed modes.

\section{Usual boundary reflection phase lags}
\label{phase lag}

In this section, we estimate in a simple way the reflection phase lags introduced near the center and surface, $\delta_{\rm s}$ and $\delta_{\rm c}$. The surface and core boundaries are supposed to be turning points denoted by $r_{\rm t}$ and such as $\mathcal{K}_r^2(r_{\rm t})=0$, at the interface between a resonant cavity and an evanescent region. Assuming the JWKB approximation is met in the adjacent resonant cavities, it is then possible to use a single turning point analysis of the stellar oscillations equations such as performed by \cite{Shibahashi1979} to deduce the wave function in the vicinity of each of these turning points. 
The analysis uses in addition the Cowling approximation, neglects the gradients in the equilibrium structure everywhere in the star and considers only regular solutions in the evanescent region (i.e., with finite values of the wave function). Within this framework, the asymptotic solution of the wave function inside the resonant cavities can be expressed and the associated reflection phase lag turns out to depend on the nature of the considered turning point.

\paragraph{In the case of a turning point such as $\sigma^2=N^2(r_{\rm t})$,}  \eq{wave equation} can be solved using a Green-Liouville transformation while neglecting the variations of the equilibrium structure (e.g., the $M$ function). The regular solution of the transformed equation  thus takes the form of an Airy function of the first kind. The asymptotic expansion of the wave function in the resonant cavity is finally given using a complex notation by \citep[e.g.,][]{Shibahashi1979,Unno1989}
\algn{
\Psi\approx\frac{a}{\sqrt{\mathcal{K}_r}} \left(\underbrace{e^{-\di \pi/4}}_{A} e^{\di |\varphi(r;r_{\rm t})|} +\underbrace{e^{\di \pi/4}}_{B} e^{-\di |\varphi(r;r_{\rm t})|} \right) e^{-\di \sigma t}\; ,
\label{asymp exp}
}
where $a$ is a complex constant, $\varphi(r;r_{\rm t})$ is defined in \eq{phase}, and we choose in this case the origin of the wave phase at the turning point $r_{\rm t}$. Since the progressive and regressive wave components have the same amplitude modulus, we first note that such a regular solution satisfies the totally reflective condition at the turning point, as assumed close to the core and surface in this work. Second, the phase lag at reflection $\delta_{\rm t}$ can be computed as the argument of the amplitude ratio of the upward to the downward energy ray when the resonant cavity is above the turning point (i.e., $r>r_{\rm t}$). When the resonant cavity is below the turning point (i.e., $r<r_{\rm t}$), the phase lag is equal to the argument of the amplitude ratio of the downward to the upward energy ray. In both cases, the phase lag at reflection is reduced to
\algn{
\delta_{\rm t}={\rm sgn}\left(\sigma^2-S_\ell^2 \right) \arg \left( \frac{A}{B}\right)={\rm sgn}\left(S_\ell^2-\sigma^2 \right) \frac{\pi}{2} \; ,
\label{delta_t 1}
}
where the first sign factor estimated well inside the cavity accounts for the case of gravity-dominated waves (i.e., $\sigma^2\ll S_\ell^2$) for which the group velocity is in the opposite direction of the phase velocity.

\paragraph{In the case of a turning point such as $\sigma^2=S_\ell^2(r_{\rm t})$,} \eq{wave equation} is singular at $r_{\rm t}$ so that we first have to solve \eq{wave equation 2} for the dependent variable $\widetilde{\Psi}$. According to \cite{Shibahashi1979}, the asymptotic expansion of $\widetilde{\Psi}$ in the resonant cavity is also given by \eq{asymp exp}. To compute the wave function $\Psi$ (which is the reference dependent variable we consider in this paper), we can use \eq{Psi by Psi_tilde} while neglecting the derivative of the $Q$ function related to the variations of the equilibrium structure. Considering the fact that \smash{$(\dd |\varphi|/\dd r)={\rm sgn}(r-r_{\rm t})~|\mathcal{K}_r|$}, we obtain:
\algn{
\Psi\approx\frac{\di a }{\sqrt{\mathcal{K}_r}} {\rm sgn}\left[\frac{\sigma^2-N^2}{ r-r_{\rm t}} \right]\left(\underbrace{e^{-\di \pi/4}}_{A^\prime} e^{\di |\varphi(r;r_{\rm t})|} \underbrace{-e^{\di \pi/4}}_{B^\prime} e^{-\di |\varphi(r;r_{\rm t})|} \right) e^{-\di \sigma t}\; ,
\label{asymp exp 2}
}
Similarly to the previous case we describe above, we thus find that the phase lag at reflection is reduced to:  
\algn{
\delta_{\rm t}={\rm sgn}\left(\sigma^2-S_\ell^2 \right) \arg \left( \frac{A^\prime}{B^\prime}\right)={\rm sgn}\left(\sigma^2-S_\ell^2 \right) \frac{\pi}{2} \; .
\label{delta_t 2}
}

\paragraph{In terms of synthesis,} the comparison between \eqsduo{delta_t 1}{delta_t 2}~shows that the change of nature in the turning point is associated with a shift in the phase lag at reflection of $\pi$ (modulo $2\pi$). Moreover, considering a mode with one single cavity located between two turning points $r_1^-$ and $r_1^+$ at the core and the surface, we see that $\delta_{\rm c}+\delta_{\rm s}=\pi$ (modulo $2\pi$) if the turning points have the same nature, namely, if $\sigma^2=N^2(r_1^-)=N^2(r_1^+)$ or $\sigma^2=S_\ell^2(r_1^-)=S_\ell^2(r_1^+)$; whereas it is $\delta_{\rm c}+\delta_{\rm s}=0$ otherwise.
\end{appendix}
\end{document}